\documentclass[12pt]{article}

\usepackage{graphicx}
\usepackage{fancybox}
\usepackage{amsmath}
\usepackage{amssymb}
\usepackage{latexsym}
\usepackage{epsfig}

\newcommand{\al}{\alpha}
\newcommand{\ep}{\epsilon}

\newcommand{\la}{\lambda}
\newcommand{\La}{\Lambda}

\newcommand{\co}{{\scriptstyle \circ}}

\newcommand{\msc}[1]{\mbox{\scriptsize #1}}
\newcommand{\dsp}{\displaystyle}

\newcommand{\br}{\Bbb R}
\newcommand{\bz}{\Bbb Z}
\newcommand{\bq}{\Bbb Q}

\newcommand{\bh}{{\Bbb H}}

\newcommand{\bsz}{\Bbb Z}

\newcommand{\cT}{{\cal T}}

\newcommand{\cN}{{\cal N}}

\newcommand{\cP}{{\cal P}}

\newcommand{\cH}{{\cal H}}

\newcommand{\cI}{{\cal I}}

\newcommand{\tL}{\tilde{L}}

\newcommand{\tj}{\tilde{j}}

\newcommand{\tell}{\tilde{\ell}}
\newcommand{\tal}{\tilde{\al}}

\newcommand{\ttau}{\widetilde{\tau}}
\newcommand{\ta}{\tilde{a}}
\newcommand{\tb}{\tilde{b}}

\newcommand{\any}{{}^{\forall}}

\renewcommand{\mod}{\, \mbox{mod} ~ }

\newcommand{\wS}{\widehat{S}}

\newcommand{\ket}[1]{{\left|#1\right\rangle}}
\newcommand{\bra}[1]{{\left\langle#1\right|}}

\newcommand{\Th}[2]{\Theta_{#1,#2}}
\renewcommand{\th}{{\theta}}

\newcommand{\tr}{\mbox{Tr}}
\renewcommand{\mod}{\mbox{mod}}

\renewcommand{\Im}{\mbox{Im}}

\newcommand{\nn}{\nonumber\\}

\newcommand{\bP}{\mbox{\bf P}}
\newcommand{\btP}{\widetilde{\mbox{\bf P}}}
\newcommand{\bD}{\mbox{\bf D}}
\newcommand{\bZ}{\mbox{\bf Z}}

\newcommand{\hZ}{\widehat{Z}}
\newcommand{\Zf}{Z^{\msc{fiber}}}
\newcommand{\hZf}{\widehat{Z^{\msc{fiber}}}}
\newcommand{\Zb}{Z^{\msc{base}}}

\newcommand{\Rf}{R_{\msc{f}}}
\newcommand{\Rb}{R_{\msc{b}}}

\def\boxit#1{\vbox{\hrule\hbox{\vrule\kern8pt
\vbox{\hbox{\kern8pt}\hbox{\vbox{#1}}\hbox{\kern8pt}}
\kern8pt\vrule}\hrule}}
\def\mathboxit#1{\vbox{\hrule\hbox{\vrule\kern8pt\vbox{\kern8pt
\hbox{$\displaystyle #1$}\kern8pt}\kern8pt\vrule}\hrule}}

\newcommand {\eqn}[1]{(\ref{#1})}

\makeatletter
\@addtoreset{equation}{section}
\def\theequation{\thesection.\arabic{equation}}
\makeatother

\begin{document}

\begin{titlepage}
 \
 \renewcommand{\thefootnote}{\fnsymbol{footnote}}
 \font\csc=cmcsc10 scaled\magstep1
 {\baselineskip=16pt
  \hfill
 \vbox{\hbox{February, 2015}
 \hbox{UTHEP-667}
       }}

 \baselineskip=20pt
\vskip 1cm
 
\begin{center}

{\bf \Large  Non-geometric Backgrounds Based on  
Topological Interfaces

} 

 \vskip 1.2cm

\noindent{ \large Yuji Satoh}\footnote{\sf ysatoh@het.ph.tsukuba.ac.jp}
 \\

\medskip

{\it Institute of Physics, University of Tsukuba, \\
 Ibaraki 305-8571,  Japan}

\vskip 8mm
\noindent{ \large Yuji Sugawara}\footnote{\sf ysugawa@se.ritsumei.ac.jp}
\\

\medskip

 {\it Department of Physical Sciences, 
 College of Science and Engineering, \\ 
Ritsumeikan University,  
Shiga 525-8577, Japan}

\end{center}

\bigskip

\begin{abstract}

We study simple models of the world-sheet CFTs describing non-geometric 
backgrounds based on the topological interfaces,  
 the `gluing condition' of which imposes T-duality- or analogous twists.
To be more specific,  we start with the torus partition function on  
a target space $S^1 [\mbox{base}] \times (S^1\times S^1 ) [\mbox{fiber}]$
with rather general values of  radii.  The fiber CFT is  defined by inserting  
the twist operators consisting of the topological interfaces which lie  
along the cycles of the world-sheet torus according to the winding numbers 
of the base circle.  We construct the partition functions 
involving such duality twists. The modular invariance is achieved straightforwardly, 
whereas  `unitarization' is  generically necessary to maintain the unitarity.
We demonstrate it in the case of the equal fiber radii.
The resultant models are closely related to the  CFTs with 
the discrete torsion.
The unitarization is  also physically 
interpreted as {\em multiple\/} insertions of the twist/interface operators 
along various directions.

\end{abstract}

\setcounter{footnote}{0}
\renewcommand{\thefootnote}{\arabic{footnote}}

\end{titlepage}

\baselineskip 18pt

\vskip2cm 
\newpage


\section{Introduction}


String theory admits the backgrounds which are non-geometric.
They are rather ubiquitous, and constitute key ingredients in understanding
intrinsic nature of quantum strings, such as string vacua and dualities.
Of particular interest among them 
are the backgrounds whose transition functions involve 
duality transformations \cite{Dabholkar:2002sy,Hellerman:2002ax,Flournoy:2004vn}. 
Since the dualities relate different backgrounds, 
the notion of the Riemannian geometry is generally lost.
In the case of T-duality, they thus result in `T-folds' \cite{Hull:2004in}.
One may also consider the backgrounds with the fluxes which are not obtained by 
a geometric compactification from a higher dimensional theory,
namely, non-geometric fluxes. 
These are often related to the geometric ones by dualities, but 
are not in general \cite{Shelton:2005cf,Shelton:2006fd,Dabholkar:2005ve}.

Beyond the classical level, the non-geometric backgrounds should be described by 
the world-sheet conformal field theory (CFT). 
The asymmetric orbifold CFTs \cite{Narain:1986qm} provide an important class,
where the left- and right-movers of the string feel different geometries.
The non-geometric backgrounds involving the duality twist generally
lie at  the fixed points of moduli, and 
the CFT description there is 
expected to be given  by certain asymmetric orbifolds
 \cite{Dabholkar:2002sy}. 
Along this line,  the T-folds with or without non-geometric fluxes 
have been studied by means of the world-sheet approaches,  {\em e.g.\/},  in
 \cite{Flournoy:2005xe,Hellerman:2006tx,Kawai:2007qd,
 Condeescu:2012sp,Condeescu:2013yma}.

In this paper, we take a step forward to study the non-geometric backgrounds
and string vacua
from the point of view of the exact world-sheet CFT.
In particular, we discuss  
them based on  the world-sheet
{\em conformal interfaces\/} \cite{Wong:1994np,Petkova:2000ip,Bachas:2001vj}.  
The conformal interfaces are defined as the one-dimensional
defects which preserve the world-sheet conformal symmetry.
When one side of the interface is empty, it becomes a conformal boundary.
In this sense, the conformal interfaces are regarded as a generalization of the conformal
boundaries, which  
describe the D-branes in string theory. 
In addition, when the left and right energy momentum tensors are separately 
continuous across the interfaces, they are called {\em topological\/} 
\cite{Petkova:2000ip,Bachas:2004sy}, since they can be deformed
freely on the world-sheet. 

The conformal interfaces possess interesting properties. 
For example, the topological interfaces implement  the symmetries and dualities of the CFT, 
including T-duality \cite{Frohlich:2004ef,Frohlich:2006ch}. 
They thus glue the CFTs which are related to each other by such symmetries or dualities.
From the target-space point of view, they may be interpreted as  submanifolds 
in a doubled target-space (bi-brane) \cite{Fuchs:2007fw}.
They also induce transformations of D-branes or boundary renormalization 
group (RG) flows \cite{Graham:2003nc,Bachas:2004sy}. 
Bulk RG flows can be described by the non-topological interfaces
\cite{Brunner:2007ur,Gaiotto:2012np}. The fusion of the conformal
interfaces is expected to give a solution-generating algebra
in string theory, similarly to the Ehlers-Geroch transformation in general relativity
\cite{Bachas:2007td}.

Since the conformal invariance is a guiding principle of the world-sheet
description of string theory, 
the conformal interface should also be a fundamental
object, though its role in string theory is yet to be uncovered.
As an attempt in this direction, we shall investigate the modular invariants 
involving  the conformal interfaces. 
Our discussion below is based on an observation that
the topological interfaces induce the twists associated with 
the symmetries and dualities,
which may yield an exact CFT description 
of T-folds and analogous non-geometric backgrounds.
We shall see that this is indeed the case by explicitly 
constructing a novel type of the modular invariants
describing such non-geometric CFT models.
For the discussions on the world-sheet conformal interfaces 
in the context of string theory, see for instance 
\cite{Bachas:2001vj,Bachas:2007td,Satoh:2011an,Bachas:2012bj,Elitzur:2013ut}.


More specifically,
we shall focus on the CFT models defined on the background, 
\begin{equation} 
S^1 [\mbox{base}] \times (S^1\times S^1) [\mbox{fiber}], 
\end{equation}  
in which the topological interfaces act on the `fiber CFT' 
when the world-sheet torus wraps around the base circle. 
In other words,  we begin with the background, 
\begin{equation} 
\br[\mbox{base}] \times \left( S^1 \times S^1 \right) [\mbox{fiber}],
\end{equation} 
and perform the `twisted compactification' implemented by the operator 
given by
\begin{equation} 
\cT_{2\pi \Rb} \otimes \cI_{\msc{fiber}}. 
\end{equation}  
Here, $\cT_{2\pi \Rb}$ denotes the translation along the base-direction, 
$\cT_{2\pi \Rb} :  X \, \mapsto \, X + 2\pi \Rb $, whereas  
$\cI_{\msc{fiber}}$ expresses a certain  
`twist' operator acting on the fiber sector, 
which is made up of the topological interfaces.
Such a twist combined with the shift in the base  
of course follows the spirit of 
the Scherk-Schwarz compactification \cite{SS1,SS2}.
However, the operator $\cI_{\msc{fiber}}$ here does not necessarily 
correspond to a symmetry of the original CFT. This means that 
the conformal systems we 
propose do not always reduce to ordinary orbifold CFTs.
In this sense,  our present study would shed light on 
a possibility to construct a novel type of string vacua,
and would illustrate a role of the world-sheet conformal interfaces 
in string theory.

~


This paper is organized as follows:

After providing the necessary notation and making a brief review 
on the topological interfaces in section 2, 
we shall start our construction of the conformal models 
and partition functions describing 
non-geometric backgrounds in section 3. We define the relevant models
by using the twisted compactification mentioned above. 
Even though the construction  would be natural and  the modular invariance is achieved, 
an issue  of unitarity remains except for the simplest case of the T-fold.


Therefore, 
in section 4, we discuss a `unitarization' of the models of our interest. 
The proposed models are closely related to the orbifold CFTs with 
the {\em discrete torsion\/} \cite{dtorsion},
and we explicitly demonstrate that they are indeed unitary for the equal fiber-radii. 
We further discuss how our unitarization is physically interpreted 
as the {\em multiple\/} insertions of the twist operators consisting of 
the topological interfaces.

In section 5, we present a summary and discussion.

~


\section{Preliminaries}

Before presenting our main analysis, we first set up the necessary notation. 
Through this paper we shall use the $\al'=1$ convention.
We set $\La \equiv \bz \tau+ \bz$, where $\tau \in \bh$ (upper half plane) is the modulus of 
the world-sheet torus parametrized as $\tau = \tau_1 + i \tau_2$ $(\tau_1 \in \br, \ \tau_2 >0)$.

~

\subsection{Partition Functions of Compact Bosons}

The partition function of a free  boson compactified on the circle with radius
$R$ should be
\begin{eqnarray}
 Z_{R} (\tau) & = & 
\sum_{\nu \in \La}\, Z_R(\tau \,|\, \nu),
\label{Z R}
\\
 Z_{R}(\tau \,|\, \nu) &: =&  \frac{R}{\sqrt{\tau_2}\,| \, \eta(\tau)|^2} 
e^{-\frac{\pi R^2}{\tau_2}|\nu|^2}, 
\label{Z nu}
\end{eqnarray} 
where $Z_{R}(\tau \,|\, \nu)$ represents a contribution from 
the winding sector specified by $\nu$.
Its modular property is expressed as 
\begin{equation}
Z_{R}(\tau+ 1 \,|\, \nu) = Z_{R}(\tau \, |  \, \nu),
\hspace{1cm} Z_{R}\left( \left. - \frac{1}{\tau}  \, \right| \, \frac{\nu}{\tau} \right)
= Z_{R}(\tau \, | \, \nu).
\label{modular Z nu}
\end{equation}

When the radius $R$ can be written as $R = \sqrt{k}$,  $(k\in \bz_{>0})$, 
the partition function $Z_R(\tau)$ is rewritten in terms of theta functions,%
\footnote
  {Our conventions of theta functions are summarized in Appendix A.} 
\begin{equation}
Z_R(\tau) =  \frac{1}{\left|\eta(\tau)\right|^2}\, \sum_{m \in \bz_{2k}}\,
\left|\Th{m}{k}(\tau) \right|^2.
\end{equation}
More generally, if $R= \sqrt{\frac{L}{K}}$, $(K,L \in \bz_{>0})$ holds (the rational CFT cases),
we obtain
\begin{equation}
Z_R(\tau) = \sum_{\stackrel{r \in \bz_{2L}}{s \in \bz_{K}}}\, 
Z^{(KL)}[Kr, Ls] (\tau),
\label{ZRtheta}
\end{equation}
with the notation, 
\begin{eqnarray}
&& Z^{(k)}[u,v](\tau) : = 
\frac{1}{\left| \eta(\tau) \right|^2}\,
\Th{u+v}{k}(\tau) \overline{\Th{-u+v}{k}(\tau)}.
\label{Zuv}
\end{eqnarray}

~


\subsection{Orbifolding}

We introduce the operators corresponding to the following two types of orbifolding: 
\begin{description}
\item[(i) $\bz_N$-action of translation : ]


First, we define $\tau^{(N), R}_{\gamma}$ as the operator linearly acting  
on the function \eqn{Z nu} as 
\begin{equation}
\tau^{(N), R}_{\gamma} \cdot \left[ \sum_{i} \, c_i \, Z_R \left(\tau | \, \nu_i \right) \right] := 
\sum_i \, c_i \, Z_R \left(\tau  \left|   \nu_i + \frac{\gamma}{N} \right.\right),
\hspace{1cm} (\any \gamma \in \La ),
\label{Z_N orb}
\end{equation}
with arbitrary $c_i$, $\nu_i$.
%
%
Acting on $Z_R(\tau)$, the operator  $\frac{1}{N}\sum_{j\in \bz_N} \tau^{(N), R}_j$ 
implements
the projection restricting 
the Kaluza-Klein (KK) momentum  to  $n \in N \bz$ after the Poisson resummation. 
By the modular completion, we then have an identity, 
\begin{equation}
Z_{R/N} (\tau)  =  \frac{1}{N} \sum_{\gamma \in \La/N\La}\, \tau^{(N), R}_{\gamma} \cdot Z_R(\tau).
\end{equation}
When $R = \sqrt{\frac{N}{L}}$, one can express
 $Z_R(\tau)$  by theta functions 
through \eqn{ZRtheta}, where $\tau^{(N),R}_{\gamma}$ act as
\begin{equation}
\tau^{(N),R}_{L(a\tau+b)} \cdot Z^{(NL)}[u,v](\tau) =
e^{2\pi i \frac{b}{N}u}\, Z^{(NL)}[u,v+La] (\tau).
\label{tau N Zuv}
\end{equation}
We set above $\gamma = L(a\tau+b)  \in \La/N\La$ with  $a,b \in \bz_N$
by assuming $N,L$ are coprime.
This form of the action turns out to be useful for our later analysis.

%
\item[(ii) $\bz_N$-action of `dual translation' : ]

We define $\ttau^{(N),R}_{\gamma}$ as the operator 
linearly acting
on the function \eqn{Z nu} as 
\begin{equation}
\ttau^{(N),R}_{\gamma} \cdot \left[ 
\sum_i c_i \, Z_R \left(\tau | \, \nu_i \right) \right]
 := 
\sum_i \, c_i Z_R \left(\tau \, | \, \nu_i \right)\,
e^{2\pi i \frac{1}{N} \langle \nu_i, \gamma \rangle}, 
\hspace{1cm} (\any \gamma \in \La ),
\label{N-fold cover}
\end{equation}
where we introduced the symbol,
\begin{equation}
\langle \nu, \gamma \rangle := \frac{1}{\tau_2} \Im (\nu \bar{\gamma}).
\label{<>}
\end{equation}
This time, the operator  $\frac{1}{N}\sum_{j\in \bz_N} \ttau^{(N),R}_j$ 
acts on $Z_R(\tau)$ as the projection onto the states with the winding $w \in N \bz$. 
We  then have an identity, 
\begin{equation}
Z_{N R} (\tau)  =  \frac{1}{N} \sum_{\gamma \in \La/N\La}\, \ttau^{(N), R}_{\gamma} 
\cdot Z_R(\tau) .
\end{equation}
When $R = \sqrt{\frac{L}{N}}$ with $L,N$ being coprime, one can express
$Z_R(\tau)$ by theta functions.
There,
$\ttau^{(N),R}_{\gamma}$ ($\gamma = L(\ta \tau + \tb) \in \La/N\La; \ \ta, \tb \in \bz_N$) 
act  as 
\begin{equation}
\ttau^{(N),R}_{L(\ta\tau+\tb)} \cdot Z^{(NL)}[u,v](\tau) =
e^{2\pi i \frac{\tb}{N}v}\, Z^{(NL)}[u+L\ta,v] (\tau),
\label{ttau N Zuv}
\end{equation}
similarly to $\tau^{(N),R}_{\gamma}$.
We also note a schematic equivalence, 
\begin{equation}
\frac{1}{N}\sum_{\gamma \in \La/N\La} \ttau^{(N),R}_{\gamma} 
= T \, \co \,  \frac{1}{N}\sum_{\gamma \in \La/N\La} \tau^{(N), 1/R}_{\gamma}
\, \co \, T,
\end{equation}
where $T$ denotes the T-duality transformation, 
$\dsp R \, \rightarrow \, \frac{1}{R}$. 

\end{description}

We shall later use the abbreviations $\tau^{(N)}_{\gamma} \equiv \tau^{(N), R}_{\gamma}$, 
$\ttau^{(N)}_{\gamma} \equiv \ttau^{(N), R}_{\gamma}$,  as long as there is no fear of confusion. 


~


\subsection{Topological Interfaces}
\label{TopInt}

Consider two  CFTs with central charge $c=1$, which are
denoted by `$\mbox{CFT}_1$' and  `$\mbox{CFT}_2$',   and
described by 
free bosons $X_1, X_2$ compactified on the circle with radius $R_1$, $R_2$,  respectively.
We assume that there exist two positive integers $k_1$, $k_2 $ such that 
\begin{equation}
\frac{k_2 R_1 R_2}{k_1} = 1.
\label{cond I-}
\end{equation}
We denote the oscillators of the free bosons $X_i$  
as $\al^i_n$, $\tal^i_n$, and the Fock vacua 
as $\ket{n, w ; (i)}$, $\bra{n,w; (i)}$ whose left and right momenta are given by 
\begin{equation}
p^{(i)}_L = \frac{n}{R_i} + w R_i , ~~~ p^{(i)}_R = \frac{n}{R_i} - w R_i, \hspace{1cm} 
(\any n, w \in \bz).
\end{equation}

Then, one has the  topological interface operator
 $I^{(-)}_{12; (k_1,k_2)}$ that glues the world-sheet of $\mbox{CFT}_2$ 
 with that of $\mbox{CFT}_1$
 \cite{Bachas:2001vj,Bachas:2007td},%
\footnote{We simply set to zero the `moduli' parameters of the topological interface operators
through this paper. We also set to plus a possible sign which could appear in the oscillator part,
since it is absorbed by exchanging the left and right movers.
It is understood that the oscillators $\alpha_n^1, \tilde\alpha_n^1$ act on the left side of
$G^{(-)}_{12; (k_1,k_2)}$, whereas  $\alpha_n^2, \tilde\alpha_n^2$ on the right side.} 
\begin{eqnarray}
I^{(-)}_{12; (k_1,k_2)} & := & G^{(-)}_{12; (k_1,k_2)} \, \prod_{n=1}^{\infty}\, e^{\frac{1}{n} \left(\al^1_{-n} \al^2_n -
\tal^1_{-n} \tal^2_n \right)},
\label{def I-}
\\
G^{(-)}_{12; (k_1,k_2)} & := & \sqrt{k_1 k_2} \, \sum_{r,s\in \bz}\,
\ket{k_1 s, k_2 r; (1)} \, \bra{k_1 r, k_2 s ; (2)}.
\label{def G-}
\end{eqnarray}
Note that the zero-mode part $G^{(-)}_{12; (k_1,k_2)}$ in \eqn{def G-} imposes the `gluing conditions', 
\begin{equation}
p_L^{(1)} = p_L^{(2)}, \hspace{1cm} p_R^{(1)} = - p_R^{(2)},
\label{gluing G-}
\end{equation}
whereas those on the oscillator part are 
\begin{equation}
\al^1_{n} \, I^{(-)}_{12; (k_1,k_2)}  = I^{(-)}_{12; (k_1,k_2)} \, \al^2_{n}, 
\hspace{1cm}
\tal^1_{n} \, I^{(-)}_{12; (k_1,k_2)}  = - I^{(-)}_{12; (k_1,k_2)} \, \tal^2_{n}.
\label{gluing I-}
\end{equation}
Namely, $I^{(-)}$ glues two theories related by T-duality. 
We also note that when $k_1 > 1$ or $k_2 >1$, the image of $I^{(-)}_{12; (k_1,k_2)}$ 
is strictly smaller than the whole Hilbert space of $\mbox{CFT}_1$. 
In other words, 
$
I^{(-)}_{21; (k_2,k_1)} \cdot I^{(-)}_{12; (k_1,k_2)} 
$
is not equal to the identity of $\mbox{CFT}_2$, but is rather identified with the projection operator 
whose image is the invariant subspace satisfying (\ref{gluing G-}).

~


There is another type of the topological interface operator $I^{(+)}$
gluing two free boson theories.  
It is given by taking the T-duality of 
the $\mbox{CFT}_1$ in the above construction,
\begin{equation}
 S^1_{R_1} \, \rightarrow \, S^1_{\tilde{R}_1}, \hspace{1cm}
 (\tilde{R}_1 \equiv 1/R_1),
\label{Tdual}
\end{equation}
where $S^1_R$ denotes the circle with radius $R$. 
The condition \eqn{cond I-} is then replaced  with
\begin{equation}
\frac{k_2 R_2}{k_1 R_1} = 1,
\label{cond I+}
\end{equation}
and its explicit form is given by
\begin{eqnarray}
I^{(+)}_{12; (k_1,k_2)} & := & G^{(+)}_{12; (k_1,k_2)} \, \prod_{n=1}^{\infty}\, e^{\frac{1}{n} \left(\al^1_{-n} \al^2_n +
\tal^1_{-n} \tal^2_n \right)},
\label{def I+}
\\
G^{(+)}_{12; (k_1,k_2)} & := & \sqrt{k_1 k_2} \, \sum_{r,s\in \bz}\,
\ket{k_2 r, k_1 s; (1)} \, \bra{k_1 r, k_2 s ; (2)}.
\label{def G+}
\end{eqnarray}
The gluing conditions for $I^{(+)}$ are 
written as 
\begin{eqnarray}\label{glueI+}
&& p_L^{(1)} = p_L^{(2)}, \hspace{1cm} p_R^{(1)} =  p_R^{(2)},
\label{gluing G+}
\\
&& \al^1_{n} \, I^{(+)}_{12; (k_1,k_2)}  = I^{(+)}_{12; (k_1,k_2)} \, \al^2_{n}, 
\hspace{1cm}
\tal^1_{n} \, I^{(+)}_{12; (k_1,k_2)}  =  I^{(+)}_{12; (k_1,k_2)} \, \tal^2_{n}.
\label{gluing I+}
\end{eqnarray}

~

From the above gluing conditions, one readily finds that
the  left and right energy-momentum tensors are separately preserved
across the interface operators. 
Thus, $I^{(\pm)}_{12; (k_1,k_2)}$
 indeed represent topological interfaces \cite{Petkova:2000ip,Bachas:2004sy}, 
which can be deformed freely on the world-sheet. 
These are special cases of more general $\widehat{u}(1)$-preserving
conformal interfaces \cite{Bachas:2001vj,Bachas:2007td}. The general gluing conditions
are concisely written  by  $O(1,1)$ matrices. The superscripts $(\pm)$
stand for the connected components of  $O(1,1)$. The topological interfaces 
which do not preserve the $\widehat{u}(1)$-symmetries have also been discussed in
\cite{Fuchs:2007tx}.

Generally, there are two important subclasses of the topological interfaces 
\cite{Frohlich:2004ef,Frohlich:2006ch}. One is named the {\em group-like defect}. 
This class of the interfaces
implements the symmetries of the CFT. The other, which includes the former, 
is the {\em duality defect}. This class   implements the order-disorder dualities or orbifold equivalences.
In our case, $I^{(\pm)}_{12; (k_1,k_2)}$ with $k_1k_2 = 1$ is  group-like and generates 
the T-dual symmetry. All other $I^{(\pm)}_{12; (k_1,k_2)}$ are duality defects 
\cite{Fuchs:2007tx,Bachas:2012bj}. 
Accordingly, our topological interfaces
result in two different classes of the `duality' twists.


~

\section{Construction of the Models}
\label{section:ConstructionModel}

Now, let us start the construction of the models describing  
non-geometric backgrounds based on the topological interfaces.

~


\subsection{Basic Set Up}
\label{BasicSetUp}

We start with a conformal system 
consisting of free bosons whose target space is given by
\begin{equation}
[\mbox{`base'}~ \br] \times [\mbox{`fiber'} ~ 
S^1_{R_{1}} \times S^1_{R_{2}}],
\end{equation}
where  
it is assumed that 
\begin{equation}
R_1 R_2 = \frac{N}{M} , ~~  (N, M \in \bz_{>0}, ~ \gcd \{N, M\} =1).
\label{assumption R1R2} 
\end{equation}
Then, the condition \eqn{cond I-} is satisfied when 
setting 
\begin{equation}
k_1 = N, ~~~ k_2 = M.
\end{equation}

We would like to make a `twisted' compactification of this system  
implemented by the operator, 
\begin{equation}
\cT_{2\pi R_{\msc{b}}} \otimes \cI^{(-)} .
\label{def twisting}
\end{equation} 
Here, $\cT_{2\pi R_{\msc{b}}}$ denotes the shift operator acting 
on the base space as $ X \, \longmapsto \, X + 2\pi \Rb  $.
The fiber part $\cI^{(-)}$ is defined in terms of 
the topological interface operator $I^{(-)}$ by 
\begin{equation}
\cI^{(-)} := \cP \left[ I^{(-)}_{12} \otimes I^{(-)}_{21} \right].
\label{def cI}
\end{equation} 
The permutation operator  $\cP$ in the above acts as
\begin{equation}
\cP \left[\ket{\al}_1 \otimes \ket{\beta}_2 \right]
= \ket{\beta}_2 \otimes \ket{\al}_1 ,
\end{equation}
where $\ket{\al}_1$, $\ket{\beta}_2$ express arbitrary states of $\mbox{CFT}_1$ and $\mbox{CFT}_2$, respectively.  
We note that $\cI^{(-)}$ is regarded as a `self-interface' of 
the fiber CFT on $S^1_{R_1} \times S^1_{R_2} $,  and thus 
any product $\left(\cI^{(-)}\right)^m$ is well-defined on its Hilbert space.  
More explicitly, the zero-mode part of $\cI^{(-)}$ is given as 
\begin{eqnarray}
\cI^{(-)}_{(0)}
\propto  
\sum_{r,s\in \bz}\, \sum_{r',s' \in \bz}\,
\left[\ket{Nr', Ms' \,;\, (1) } 
\otimes \ket{Ns, Mr \, ; \, (2)} \right]
\, \left[\bra{Nr, Ms \,;\, (1)} 
\otimes \bra{N s', M r' \, ;\, (2)}  \right].
\label{cI-0}
\end{eqnarray}
As  mentioned above, 
the topological interface $I^{(-)}$ induces the 
duality transformations. Especially, $\cI^{(-)}_{(0)}$ glues 
the zero-mode parts
so as to interchange 
the KK momentum (winding) of the $S^1_{R_1}$-theory with winding (KK momentum) of the $S^1_{R_2}$-theory. In this way, we observe  {\em non-geometric} nature 
due to the duality twist by  \eqn{def cI}.%
\footnote
{It would be possible that the models constructed  
  here are equivalent to the world-sheet CFT
 for  some geometric orbifolds via T-duality transformations.
In fact, as discussed later, this is the case for the simplest 
case with $N=M=1$, similarly to known examples of T-fold backgrounds.   
However, in generic cases with $NM>1$, it seems hard to reinterpret our models as 
CFTs realized geometrically.  
}

Now,  our ansatz of the partition function twisted by \eqn{def twisting} 
is expressed in the form,
\begin{eqnarray}
Z(\tau) & = &  \sum_{\la \in \La \left( \equiv \bz \tau + \bz\right)}\, 
\Zb_{\la}(\tau) \, \Zf_{\la} (\tau)
\nn
& \equiv &  \sum_{\la \in \La}\,  Z_{R_{\msc{b}}}(\tau\, | \, \la) \,  
\Zf_{\la} (\tau).
\label{Z general 0}
\end{eqnarray}
Here, the base part is expanded by the contribution from each
winding sector specified by $\lambda$, 
and the fiber part is correlated according to this base winding $\lambda$.
Adopting this winding basis, the twist operator \eqn{def twisting} 
induces a shift of the winding due to $\cT_{2\pi R_{\msc{b}}}$ as well as  
 the duality twist caused by $\cI^{(-)}$.  
Every time the string wraps around the base cycle,
the fiber part thus receives an additional duality twist by $\cI^{(-)}$. 
%
%
Consequently, 
the contribution from the fiber sector with the winding
$\la = m \in \bz$ (temporal winding) is given by
\begin{eqnarray}
\Zf_{(0,m)}(\tau) \equiv \Zf_{\la=m}(\tau)  :=  \cN_\la \, \tr_{\cH_{\msc{fiber}}} 
\left[\left(\cI^{(-)}\right)^{|m|} \, q^{L_0-\frac{2}{24}}
\bar{q}^{\tL_0-\frac{2}{24}}\right],
\label{def Zf 0m}
\end{eqnarray}
where $\cN_{\la}$ is a normalization constant.

The torus partition functions with multiple insertions of the general 
$\widehat{u}(1)$-preserving conformal interfaces have been evaluated 
in \cite{Sakai-Satoh}. The evaluation of \eqn{def Zf 0m} is much simpler and,   
once it is obtained explicitly,  
we can uniquely determine the fiber partition function with general winding 
$
\Zf_{(w,m)}(\tau) \equiv \Zf_{\la=w\tau+ m}(\tau)
$
so that the total partition function \eqn{Z general 0} becomes  
modular invariant. 
This means that $\Zf_{\la} (\tau)$ should possess the modular properties, 
\begin{equation}
\Zf_{\la}(\tau+1) = \Zf_{\la}(\tau), \hspace{1cm} \Zf_{\la/\tau} \left(-\frac{1}{\tau} \right) = \Zf_{\la}(\tau).
\label{Zf modular covariance}
\end{equation}
Here, we should note that $\la$ and $\tau$ are treated as independent variables.  
One may rephrase \eqn{Zf modular covariance} in terms of the alternative notation $\Zf_{(w,m)}(\tau)$ as 
\begin{equation}
\Zf_{(w,m)}(\tau+1) = \Zf_{(w, w+m)}(\tau), \hspace{1cm} \Zf_{(w,m)} \left(-\frac{1}{\tau} \right) = \Zf_{(m,-w)}(\tau).
\label{Zf modular covariance 2}
\end{equation}
Based on these relations, 
one can readily generate 
the general building blocks $\Zf_{(w,m)}(\tau)$ from $\Zf_{(0,m)}(\tau)$.
We shall also  assume the `parity invariance',
\begin{equation}
\Zf_{-\la} (\tau) = \Zf_{\la} (\tau),
\end{equation}
which seems physically natural  and in accord with the twist operator \eqn{def twisting}.  
In particular, the normalization constants should satisfy
\begin{equation}
\cN_{w\tau+m} = \cN_{w\tau+(w+m)}=\cN_{m\tau-w}=\cN_{-(w\tau+m)}.
\end{equation}
We later choose these constants suitably.  


~

From now on,
let us analyze concrete examples.


\subsection{$N=M=1$ case}
\label{N=M=1}

We first focus on  the simplest case of $N=M=1$. 
This case also serves as a guide for the general cases that we discuss later. 
It turns out that it suffices to set $\cN_\la = 1$ in this case.
We also set $R_1 = 1/R_2 = R_{\msc{f}}$. We then
evaluate $\Zf_{(0,m)}$ in \eqn{def Zf 0m} for $m \in 2\bz+1$ and $m \in 2\bz - \{0\}$, 
separately.

~


\noindent
\underline{\bf  $m\in 2\bz+1$ sectors :}

We first consider the $m=1$ case.
For the zero-mode part, 
we obtain 
\begin{eqnarray}
\Zf_{(0,1), \msc{zero-mode}} (\tau) & = & \sum_{r,s\in \bz}\,  \sum_{r',s'\in \bz}\,
\delta_{r,r'} \delta_{s,s'}\,
q^{\frac{1}{4} \left(\frac{r}{\Rf} +  \Rf s \right)^2} 
\bar{q}^{\frac{1}{4} \left(\frac{r}{\Rf} -  \Rf s \right)^2}
q^{\frac{1}{4} \left(\frac{r'}{\Rf} +  \Rf s' \right)^2} 
\bar{q}^{\frac{1}{4} \left(\frac{r'}{\Rf} -  \Rf s' \right)^2}
\nn
& = & \sum_{r,s \in \bz} \, 
q^{\frac{1}{4} \left(\frac{2 r}{\sqrt{2} \Rf} +  \sqrt{2} \Rf s \right)^2} 
\bar{q}^{\frac{1}{4} \left(\frac{2 r}{\sqrt{2} \Rf} -  \sqrt{2} \Rf s \right)^2}.
\label{N=1 zero-modes}
\end{eqnarray}
In this calculation,  
only the `diagonal part' of $\mbox{CFT}_1 \otimes \mbox{CFT}_2$ with 
\begin{equation}
p_L^{(1)} = p_L^{(2)}= \frac{r}{\Rf} + s \Rf, \hspace{1cm} 
p_R^{(1)} = - p_R^{(2)}= \frac{r}{\Rf} - s \Rf
\end{equation}
survives in the trace under the insertion of $\cI^{(-)}$. 
This eventually leads to the zero-mode spectrum of a compact boson of  
radius $\sqrt{2} \Rf$,   where 
the KK momenta are restricted to even numbers, 
that is, $r' = 2r \in 2\bz$.

The oscillator part also only includes the diagonal part of
 $\mbox{CFT}_1 \otimes \mbox{CFT}_2$, namely, 
\begin{eqnarray}
\Zf_{(0,1), \msc{oscillator}} (\tau) & = & q^{\frac{2}{24}} \bar{q}^{\frac{2}{24}} \, 
\prod_{n, \tilde{n} =1} \, \frac{1}{1-q^{2n}} \, \frac{1}{1-\bar{q}^{2\tilde{n}}}
\nn
& = & \left|\frac{2\eta(\tau)}{\th_2(\tau)}\right| \, \frac{1}{\left|\eta(\tau)\right|^2}.
\label{N=1 osc}
\end{eqnarray}
Combining \eqn{N=1 zero-modes} and \eqn{N=1 osc}, we obtain 
\begin{eqnarray}
\Zf_{(0,1)}(\tau) &=& \left|\frac{2\eta(\tau)}{\th_2(\tau)}\right| \, \frac{1}{\left|\eta(\tau)\right|^2} \,
  \sum_{r,s \in \bz} \, 
q^{\frac{1}{4} \left(\frac{2 r}{\sqrt{2} \Rf} +  \sqrt{2} \Rf s \right)^2} 
\bar{q}^{\frac{1}{4} \left(\frac{2 r}{\sqrt{2} \Rf} -  \sqrt{2} \Rf s \right)^2}
\nn
& = & \left|\frac{2\eta(\tau)}{\th_2(\tau)}\right|\,
\frac{1}{2} \sum_{\nu \in \bz\tau+ \frac{1}{2}\bz }\, Z_{\sqrt{2}\Rf} (\tau \, | \, \nu).
\label{N=1 Zf 01}
\end{eqnarray}
In the second line we made use of the Poisson resummation. 
It is easy to confirm that we reach the same result  for any $m\in 2\bz+1$: 
\begin{eqnarray}
\Zf_{(0,m)}(\tau) &=& \left|\frac{2\eta(\tau)}{\th_2(\tau)}\right|\,
\frac{1}{2} \sum_{\nu \in \bz\tau+ \frac{1}{2}\bz }\, Z_{\sqrt{2}\Rf} (\tau \, | \, \nu), 
\hspace{1cm}
(\any m \in 2\bz+1).
\label{N=1 Zf 0 odd}
\end{eqnarray}
The interpretation of the odd sectors 
as the diagonal part becomes important in the later discussions.

~


\noindent
\underline{\bf $m\in 2\bz - \{0\}$ sectors :}

The even sectors are  easy to compute, since we have $\left( \cI^{(-)}\right)^2 = 1$ for the $N=M=1$ case. 
We then  obtain
\begin{eqnarray}
\Zf_{(0,m)}(\tau) & =& \frac{1}{\left|\eta(\tau)\right|^4}\,
\left[
\sum_{r,s \in \bz} \, 
q^{\frac{1}{4} \left(\frac{r}{\Rf} +   \Rf s \right)^2} 
\bar{q}^{\frac{1}{4} \left(\frac{r}{\Rf} -  \Rf s \right)^2}
\right]^2
\nn
& = & \left[\sum_{\nu \in \La} \, Z_{\Rf}(\tau \, | \, \nu)\right]^2
\equiv \Zf_{(0,0)}(\tau).
\label{N=1 Zf 0 even}
\end{eqnarray}

~


\noindent
\underline{\bf general winding sectors :}

We finally determine all the  winding sectors that are compatible with the modular invariance. 
Let $\la \equiv w\tau +m \in \La$ be the winding of the base circle. 
Then, it turns out that the partition functions in question are given by
\begin{itemize}
\item $\la \in 2\La $ :
\begin{eqnarray}
\Zf_{\la} (\tau) &=&  \left[\sum_{\nu \in \La} \, Z_{\Rf}(\tau \, | \, \nu)\right]^2.
\end{eqnarray}

\item $\la \in 2\La+1$ :
\begin{eqnarray}
\Zf_{\la} (\tau) 
&=& 
\left|\frac{2\eta(\tau)}{\th_2(\tau)}\right|\,
\frac{1}{2} \sum_{j \in \bz_2}\, \sum_{\nu \in \La + \frac{j}{2} }\, Z_{\sqrt{2}\Rf} (\tau \, | \, \nu).
\end{eqnarray}

\item $\la \in 2\La + \tau$ :
\begin{eqnarray}
\Zf_{\la} (\tau) &=&  
\left|\frac{2\eta(\tau)}{\th_4(\tau)}\right|\,
\frac{1}{2} \sum_{j \in \bz_2}\, \sum_{\nu \in \La + \frac{j}{2}\tau }\, 
Z_{\sqrt{2}\Rf} (\tau \, | \, \nu).
\end{eqnarray}

\item $\la \in 2\La + \tau+1 $ :
\begin{eqnarray}
\Zf_{\la} (\tau) &=&  
\left|\frac{2\eta(\tau)}{\th_3(\tau)}\right|\,
\frac{1}{2} \sum_{j \in \bz_2}\, \sum_{\nu \in \La + \frac{j}{2}(\tau+1) }\, 
Z_{\sqrt{2}\Rf} (\tau \, | \, \nu).
\end{eqnarray}

\end{itemize}
The total partition function is obtained by substituting these results into \eqn{Z general 0}.

Note that the present model would be identified with an example 
of the T-folds, that is, the non-geometric backgrounds based on the T-duality 
twists. We emphasize that this model includes a continuous modulus $\Rf$,
while
most examples in the literature, of which torus partition functions are precisely calculable, 
are well-defined only 
at special points of the moduli space (say, the self-dual radius of the circle).
Generally, the T-folds lie at the fixed points of the moduli space under the T-duality twists 
\cite{Dabholkar:2002sy}. In our setting, this is translated into the fact that the twist 
operator $\cI^{(-)}$, which is composed of the topological interface operators, 
is a `self-interface' acting
 within one Hilbert space. Indeed, the K\"ahler modulus of the $T^2$ compactification 
 is fixed under the twist induced by $\cI^{(-)}$, though the complex structure modulus is not. 

~


\subsection{General $N$, $M$}
\label{NM>1}

We next consider more general cases of 
\begin{equation}
R_1 R_2 = \frac{N}{M} , ~~~ (N, M\in \bz_{>0}), \hspace{1cm} 
k_1 = N, ~~ k_2 =M.
\label{cond general N M}
\end{equation}
It seems again the simplest to first consider the sectors with the base winding $(0,m)$.    
Relevant calculations are similar to those in the 
$N=M=1$ case, but we have a few 
differences: 
\begin{itemize}
\item For the $m \in 2\bz+1$ sectors, 
we find the zero-mode spectrum of radius $\sqrt{2 }  R_1$
with the restriction that 
$\mbox{(KK momentum)} \in 2 N  \bz$  
and $\mbox{(winding)}  \in M \bz$, 
which means
\begin{equation}
p_L= \frac{2N r}{\sqrt{2}  R_1} + \sqrt{2} M  s R_1, ~~~ 
p_R= \frac{2N r}{\sqrt{2}  R_1} - \sqrt{2} M  s R_1, ~~~ (r,s \in \bz).
\end{equation}
\item For  the  $m \in 2\bz- \{0\}$ sectors, 
we obtain the square of 
the partition function of a compact boson of 
radius $ R_1$
with the restriction
that $\mbox{(KK momentum)} \in N  \bz$ and $\mbox{(winding)}  \in M \bz$,
which means 
\begin{equation}
\hspace{-5mm}
p_L= \frac{N  r}{R_1} + M R_1 s \left( \equiv M R_2 r + \frac{N s}{R_2}\right), ~~~ 
p_R= \frac{N  r}{R_1} - M R_1 s \left( \equiv M R_2 r - \frac{N s}{R_2}\right), 
~~~ (r,s \in \bz).
\end{equation}
Note that 
\begin{equation}
\Zf_{(0,m)}(\tau) \neq \Zf_{(0,0)}(\tau), \hspace{1cm} (\any m \in 2\bz - \{0\}),
\end{equation}
contrary to the $N=M=1$ case. 
This is because $\left(\cI^{(-)}\right)^2$ acts as a projection operator 
when $k_1 >1$ or $k_2 > 1$. 
\end{itemize}


The restrictions of the zero mode spectra given above are suitably achieved by inserting 
the orbifolding operators $\tau^{(*)}_j\equiv \tau^{(*), R_i}_j$, $\ttau^{(*)}_j \equiv \ttau^{(*), R_i}_j$ 
defined in  \eqn{Z_N orb}, \eqn{N-fold cover}.
Namely, we obtain the fiber partition functions for the temporal winding sectors as: 
\begin{eqnarray}
\Zf_{(0,m)}(\tau) & =& \cN_m\,
\left|\frac{2\eta(\tau)}{\th_2(\tau)}\right| \, \frac{1}{\left|\eta(\tau)\right|^2} \,
  \sum_{r,s \in \bz} \, 
q^{\frac{1}{4} \left(\frac{2 N  r}{\sqrt{2} R_1} +  \sqrt{2} M s R_1  \right)^2} 
\bar{q}^{\frac{1}{4} \left(\frac{2 N  r}{\sqrt{2} R_1} -  \sqrt{2} M s R_1  \right)^2}
\nn
& = & \cN_m\,
 \left|\frac{2\eta(\tau)}{\th_2(\tau)}\right|\,
\frac{1}{2NM} \sum_{j \in \bz_{2N}} \, \sum_{\tj \in \bz_M} \,  
\tau_j^{(2N)} \, \ttau_{\tj}^{(M)} \cdot
Z_{\sqrt{2}  R_i} (\tau)
\nn
&& 
\hspace{8cm}
(\any m \in 2\bz+1).
\label{N M Zf 0 odd}
\\
\Zf_{(0,m)}(\tau) 
& =& \cN_m\,
\frac{1}{\left|\eta(\tau)\right|^4}\,
\left[
\sum_{r,s \in \bz} \, 
q^{\frac{1}{4} \left(\frac{N r}{R_1} +   M s R_1  \right)^2} 
\bar{q}^{\frac{1}{4} \left(\frac{N r}{ R_1} -  M s R_1  \right)^2}
\right]^2
\nn
& =&  \cN_m\,
\left[\frac{1}{N M } \sum_{j\in \bz_{N}} \, \sum_{\tj \in \bz_M} \,
\tau_j^{(N)} \, \ttau_{\tj}^{(M)} \cdot 
Z_{R_i} (\tau ) \right]^2
\nn
&& 
\hspace{8cm} (\any m \in 2\bz -\{0\}).
\label{N M Zf 0 even}
\end{eqnarray}
Here, $i =$ 1 or 2. Both expressions are the same 
due to 
the relation \eqn{cond general N M}. This also implies that the result in \eqn{N M Zf 0 even}
is  rewritten in a symmetric form with respect to $R_1$ and $R_2$. 
Moreover, we should note that the normalization constants of the interface operators 
appearing in \eqn{def G-} have been absorbed into the redefinition of $\cN_{m}$
in these expressions \eqn{N M Zf 0 odd}, \eqn{N M Zf 0 even}.
For the special case of $m=0$, we of course obtain 
\begin{equation}
\Zf_{(0,0)} (\tau) = \cN_0\,  Z_{R_1}(\tau) Z_{R_2}(\tau).
\label{N M Zf 00}
\end{equation}

The total partition function is now determined  by the modular invariance.
For notational convenience, 
we introduce, 
\begin{equation}
g_{\la}(\tau) := 
\left\{
\begin{array}{ll}
\vspace{2mm}
\left|\frac{2\eta(\tau)}{\th_2(\tau)}\right|,
& ~~ (\la \equiv 1 ~ \mod\  2\La),
\\
\vspace{2mm}
\left|\frac{2\eta(\tau)}{\th_4(\tau)}\right|,
& ~~ (\la \equiv \tau ~ \mod\  2\La),
\\
\left|\frac{2\eta(\tau)}{\th_3(\tau)}\right|,
& ~~ (\la \equiv \tau+ 1 ~ \mod\  2\La),
\end{array}
\right.
\label{def g la}
\end{equation}
for $\any \la \in \La$.
We also define the following set $S^{(K)}[\la] \subset \La/K \La$ for $\any \la \in \La$, $\any K \in \bz_{>0}$ by 
\begin{equation}
S^{(K)}[\la] := \{ \nu \in \La\,;\, \langle \la, \nu \rangle =0 \}/K \La,
\label{SK la}
\end{equation}
where the symbol $\langle ~ , ~ \rangle $ is defined in \eqn{<>}.


With these preparations, we can write down the partition functions 
of the fiber CFT in general winding sectors as follows:
\begin{description}
\item[\underline{even sectors : $(\la \in 2 \La -\{0\})$}]  
\begin{eqnarray}
 \Zf_{\la} (\tau) 
 &=& \cN_{\la}\, 
\left[\frac{1}{NM} \sum_{\al \in S^{(N)}[\la]}\,  \sum_{\tal \in S^{(M)}[\la] } \,
\tau_{\al}^{(N)} \, \ttau_{\tal}^{(M)} \cdot 
Z_{ R_i} (\tau ) \right]^2.
 \label{N M Zf even}
\end{eqnarray} 

\item[\underline{odd sectors : $(\la \in \La - 2 \La)$ }]
\begin{eqnarray}
\Zf_{\la} (\tau) &=&  \cN_{\la}\,  g_{\la}(\tau) \,
\frac{1}{2NM} \sum_{\al \in S^{(2N)}[\la]}\,  \sum_{\tal \in S^{(M)}[\la]} \, 
\tau_{\al}^{(2N)} \, \ttau_{\tal }^{(M)} \cdot
Z_{\sqrt{2} R_i} (\tau). 
\label{N M Zf odd}
\end{eqnarray}
\end{description}
Here,  $i = 1$ or $2$. 
For $\la=0$, the fiber partition function is given by \eqn{N M Zf 00}.

Note that 
\begin{equation}
S^{(N)}[\la] = \{\nu = 0\tau+j\,;\, j\in \bz\}/N\La \cong \bz_N,
\end{equation}
for the temporal winding sectors $\la = m \,  (\neq 0) \in \bz $, 
and thus \eqn{N M Zf even}, \eqn{N M Zf odd} reduce to the previous ones 
\eqn{N M Zf 0 even}, \eqn{N M Zf 0 odd}.
Due to the $SL(2; \bz)$-invariance of $\langle ~, ~ \rangle$ in \eqn{<>}, 
they also possess the expected modularity
\eqn{Zf modular covariance} or \eqn{Zf modular covariance 2}.
The commutativity of $\tau_{\al}^{(2N)} $ and $\ttau_{\tal }^{(M)}$ for general
$\la$ follows from that for $\la = m$.


We next determine the normalization factors $\cN_{\la}$ for $\any \la \in \La$.
We shall choose these constants to be the smallest positive numbers 
such that the $q$-expansion of $\Zf_{\la} (\tau)$
is written in the form, 
\begin{equation}
\Zf_{\la} (\tau) =  \sum_{\ell, \tell}\, \sum_{n, \tilde{n} \in \bz} \, a (\ell,\tell, n, \tilde{n} ; \la) \, 
q^{\varDelta(\ell)+n}\, \overline{q^{\widetilde{\varDelta}(\tell) + \tilde{n}}},
\hspace{1cm} \left|a (\ell,\tell, n, \tilde{n} ; \la)\right| \in \bz_{\geq 0},
\label{wanted q-exp}
\end{equation}
for every $\la $. 
We note, for instance, the  following $q$-expansion 
of the function $(\any \ell \in \bz- \{ 0\})$, 
\begin{equation}
\frac{1}{NM} \sum_{\al \in S^{(N)}[\ell \tau]}\,  \sum_{\tal \in S^{(M)}[\ell \tau] } \,
\tau_{\al}^{(N)} \, \ttau_{\tal}^{(M)} \cdot 
Z_{ R_1} (\tau ) 
\equiv 
\frac{1}{NM} \sum_{j \in \bz_N }\,  \sum_{\tj \in \bz_M } \,
\tau_{j \tau}^{(N)} \, \ttau_{\tj \tau}^{(M)} \cdot 
Z_{ R_1} (\tau) ,
\end{equation}
always includes  fractional coefficients taking the values in $\frac{1}{NM} \bz$ up to some phase factors.
We thus simply set
\begin{equation}
\cN_{\la} := \left\{
\begin{array}{ll}
(NM)^2 & ~~ \la \in 2\La, \\
NM & ~~ \la \in \La- 2\La.
\end{array}
\right.
\label{cN la}
\end{equation}
It is easy to confirm that, with this choice of $\cN_{\la}$,
the fiber partition function 
$\Zf_{\la}(\tau)$ is actually written in the form \eqn{wanted q-exp}.
We later discuss the validity of these normalization constants \eqn{cN la},
after introducing the `unitarized  model'  in section 4. 
The above choice is also regarded as specifying the weights 
of the summation over the winding sectors, which is equivalent 
to that over the interfaces along various cycles,  
as discussed shortly in the next subsection.


The total partition function is 
obtained by substituting these results \eqn{N M Zf even}, \eqn{N M Zf odd} and 
\eqn{cN la} into \eqn{Z general 0}.

~


Let us finally comment 
on  a useful rewriting of the relevant partition functions. 
If recalling the analysis for the simplest case $N=M=1$, the partition functions
for the odd sectors are expected to be interpreted as the `diagonal parts'
of the even sectors.  
We here demonstrate  this is indeed the case.
Set 
\begin{eqnarray}
Z^{(0)}_{\la}(\tau) & := & \sum_{\al \in S^{(N)}[\la]}\,  \sum_{\tal \in S^{(M)}[\la] } \,
\tau_{\al}^{(N)} \, \ttau_{\tal}^{(M)} \cdot 
Z_{ R_1} (\tau ),
\label{Z(0)}
\end{eqnarray}
for $\any \la \in \La  $. In terms of this, 
\begin{equation}
\Zf_{\la}(\tau) = Z^{(0)}_{\la}(\tau)^2,
\label{Zfeven}
\end{equation}
for the even sector $\any \la \in 2\La -\{0\}$. 
Moreover, the partition function \eqn{N M Zf 0 odd} for the sector $\la = m \in 2\bz+1$  
with \eqn{cN la} is rewritten as 
\begin{equation}
\Zf_{m} (\tau) = \bD_{[m]} \cdot \left[ Z^{(0)}_{2 m}(\tau)^2 \right] \equiv 
Z^{(0)}_{2m} (2\tau),
\label{rel Zf m bD}
\end{equation}
where $\bD_{[\la]}$ denotes the operator 
extracting the diagonal part defined in 
\eqn{def bDwm} or \eqn{def bDwm 2}
($[\la] \equiv \la  ~ \mod\, 2\La$). 
It is a straightforward task to define the `diagonal part operator' 
$\bD_{[\la]}$ ($\la \in \La-2\La$)
to preserve the modular covariance, as is illustrated in Appendix B. 
Thus, by the modular transformation, 
\eqn{rel Zf m bD} extends to the case with general $\la \in \La - 2 \La$:
\begin{equation}
\Zf_{\la} (\tau) = \bD_{[\la]} \cdot \left[ Z^{(0)}_{2 \la}(\tau)^2 \right].
\label{rel Zf la bD}
\end{equation}

In this way we can write down the total 
partition function in a compact form%
\footnote{Since $S^{(K)}[\ell \la] = S^{(K)}[\la]$ ($\any \ell\neq 0$), one may replace 
   $Z^{(0)}_{2 \la}(\tau)$ in \eqn{Ztot N M} with $Z^{(0)}_{\la}(\tau)$.
   The above notation, however, makes it clear that 
   the fiber partition functions for the odd sectors are regarded as
   the diagonal part of the  even sectors. 
   The `unitarized model' discussed in section 4 is defined 
   through $\wS^{(K)}[\la]$ in \eqn{hat SK la},
   instead of $S^{(K)}[\la]$, and 
    $\widehat{Z}^{(0)}_{2 \la}(\tau)$    is  distinguished from $\widehat{Z}^{(0)}_{\la}(\tau)$
    since $\wS^{(K)}[2\la] \neq  \wS^{(K)}[\la]$ generically.
} 
\begin{eqnarray}
Z(\tau) 
& = & (NM)^2\, \Zb_{0}(\tau) \, Z_{R_1}(\tau) Z_{R_2}(\tau) 
+ \sum_{\la \in 2\La-\{0\}}\, \Zb_{\la}(\tau) \, Z^{(0)}_{\la}(\tau)^2 
\nn
&&
\hspace{2cm}
+ \sum_{\la \in \La- 2\La} \, \Zb_{\la}(\tau) \, \bD_{[\la]} \cdot 
\left[ Z^{(0)}_{2 \la}(\tau)^2 \right].
\label{Ztot N M}
\end{eqnarray}
It is also easy to confirm that  \eqn{Ztot N M}  actually reduces to the total partition function given in 
section \ref{N=M=1}, when setting $N=M=1$.

~


\subsection{Interface Operator $\cI^{(-)}$ along Various Cycles}
\label{various}

In our argument so far, $\la = w \tau + m \in \La$ has 
represented the winding 
along the base circle.%
\footnote{When 
$\la = w \tau + m \in \La$ represents the winding of the base circle,
$w$ and $m$  are the spatial and the temporal windings, respectively.
They are
related to the boundary condition of the base boson as
$
X(z+2\pi i \gamma, \bar{z}- 2\pi i \bar{\gamma}) = X(z,\bar{z}) + 2\pi \Rb \langle \la, 
\gamma \rangle
$
for any cycle $\gamma \in \La$. 
}
From the point of view of the fiber sector, which is correlated to the base sector
according to $\la$, one may also regard it as specifying 
the homology cycles 
of the world-sheet torus along which the interface lies. 
This would become evident, once we take the standard parameterization of the homology 
cycles so that they  are labeled by 
$\gamma \in \La (\equiv \bz \tau + \bz)$ and 
the  `basic spatial (temporal) cycle' 
is associated with $\gamma = 1$ $(\gamma= \tau)$.

To be more specific, let us consider the interface along the cycle $\gamma \in \La$
created by the operator $\cI^{(-)}$ in \eqn{def cI}. 
We denote the corresponding fiber partition function with this interface inserted by 
`$\Zf[\cI^{(-)}_{\gamma}] (\tau)$'.
It is quite natural to 
interpret $\Zf_{\la}(\tau)$ given in \eqn{N M Zf even} and \eqn{N M Zf odd} to be 
this partition function for $\ \gamma = \la  \in \La$ up to the normalization 
factor $\cN_{\la}$;  namely,
\begin{equation}
\Zf_{\la}(\tau) \equiv \cN_{\la} \, \Zf[\cI^{(-)}_{\la}] (\tau).
\label{def ZcI}
\end{equation}
Indeed, for  $\lambda = m \in \bz$, the left side reduces to \eqn{def Zf 0m}, which 
represents the $m$-fold insertion of the interface operator along the basic spatial cycle.
By the modular transformation, such a picture extends to the case of general $\lambda$.   
We take \eqn{def ZcI} to define the homology cycles for the interfaces in the following 
discussion.

With additional notation, the fiber partition function $\Zf[\cI^{(-)}_{\la}] (\tau)$
is written in a concise form.
To this end, 
let us first introduce the subset $\La_0 \subset \La$ defined by 
\begin{equation}
\La_0 := \left\{ 1, \tau\right\} \cup \left\{ n_1 \tau + n_2\, ; \, n_1 \in \bz_{>0}, ~ n_2 \in \bz -\{0\}, 
~ \gcd \{n_1, |n_2| \} =1
\right\}.
\label{La0}
\end{equation}
Then, $\any \gamma \in \La$ can be uniquely expressible as 
$\gamma = n_{\gamma} \gamma_0$, $n_{\gamma} \in \bz$, $\gamma_0 \in \La_0$.
We also introduce the `polarized projection operators', 
\begin{equation}
\bP^{(N)}_{\la} := \frac{1}{N} \sum_{\al \in S^{(N)}[\la]}  \, \tau_{\al}^{(N)},
\hspace{1cm}  \btP^{(M)}_{\la} := \frac{1}{M} \sum_{\tal \in S^{(M)}[\la]}\, \ttau_{\tal}^{(M)} ,
\label{P la}
\end{equation}
where $S^{(K)}[\la]$ is defined in \eqn{SK la}.
Note that we can rewrite \eqn{P la} as 
\begin{equation}
\bP^{(N)}_{\la} = 
\frac{1}{N} \sum_{\al \in S^{(N)}[\la]}  \, \tau_{M \al}^{(N)},
\hspace{1cm}  \btP^{(M)}_{\la} 
 = \frac{1}{M} \sum_{\tal \in S^{(M)}[\la]}\, \ttau_{N \tal}^{(M)} ,
\end{equation}
since $N$ and $M$ are coprime.
This fact implies that the operators 
$\bP^{(N)}_{\la} $, $\btP^{(M)}_{\la'} $ commute with each other for $\any \la, \la' \in \La$,
which follows from the definitions of $ \tau_{\gamma}^{(N)}$, 
$ \ttau_{\gamma}^{(N)}$ in \eqn{Z_N orb}, \eqn{N-fold cover}.
The following identities are also obvious by definition, 
\begin{equation}
\bP^{(N)}_{n \la} = \bP^{(N)}_{\la}, \hspace{1cm} \btP^{(N)}_{n \la} = \btP^{(N)}_{\la}, 
\label{Pnla}
\end{equation}
for $\any n \in \bz$.

In terms of these polarized projection operators, 
\eqn{N M Zf even}, \eqn{N M Zf odd} and \eqn{rel Zf la bD}
are rephrased as follows: 
\begin{description}
\item[\underline{even sectors : ($n_{\gamma} \in 2\bz - \{0\}$)}]

\begin{equation}
\Zf[\cI^{(-)}_{\gamma}] (\tau) = \Zf[\cI^{(-)}_{2 \gamma_0}] (\tau)
\equiv 
\left[\bP^{(N)}_{\gamma_0} \btP^{(M)}_{\gamma_0} \cdot Z_{R_1} (\tau) \right]^2.
\label{Zf cI even}
\end{equation}

\item[\underline{odd sectors : ($n_{\gamma} \in 2\bz+1$)}]

\begin{equation}
\Zf[\cI^{(-)}_{\gamma}] (\tau) = \Zf[\cI^{(-)}_{\gamma_0}] (\tau)
\equiv   \bD_{[\gamma]} \cdot  \Zf[\cI^{(-)}_{ 2 \gamma} ] (\tau).
\label{Zf cI odd}
\end{equation}
\end{description}
As in the case of $\la = m \in \bz$, 
the  relevant cycles  reduce down to $2\gamma_0$ or $\gamma_0$,
even when $\gamma$ is a multiple of them.
By the definition \eqn{def ZcI}, `$\cI^{(-)}_{2\gamma_0}$'
is also equivalent to the two-fold insertion of $\cI^{(-)}_{\gamma_0}$
along the `short cycle' $\gamma_0$.
These expressions are generalized  later in section \ref{multiple}.

~


\subsection{Relation to the `$I^{(+)}$-twisted' Models}
\label{I+}

It would be a natural question what happens if we consider the models twisted 
by $I^{(+)}$ in place of $I^{(-)}$.
As mentioned in section \ref{TopInt}, 
$I^{(+)}$ is obtained by taking the T-duality for the $\mbox{CFT}_1$  
as in \eqn{Tdual}.
This means that
\begin{eqnarray*}
&& [\mbox{$I^{(+)}$-model for the $S^1_{\tilde{R}_1} \times S^1_{R_2}$-fiber}] \, 
\stackrel{\msc{T-duality}}{\cong} \,
[\mbox{$I^{(-)}$-model for the $S^1_{R_1} \times S^1_{R_2}$-fiber}],
\end{eqnarray*} 
with $\tilde{R}_1 = 1/R_1$ and $k_1=N$, $k_2=M$.

In the special case with $\tilde{R}_1 = 1/R_1 = R_2 $ and $k_1=k_2 = 1$, 
the interface operator $\cI^{(+)}$ composed of $I^{(+)}$   becomes 
the permutation operator $\cP$.
Therefore, 
the $I^{(-)}$-model for the $S^1_{1/R_2} \times S^1_{R_2}$-fiber
can be interpreted as  the T-dual of the `permutation twisted model',
that is, 
the orbifold of $S^1_{\Rb} \times \left[ S^1_{R_2} \times S^1_{R_2} \right]$  
twisted by  
\begin{equation}
\cT_{2\pi \Rb} \otimes \cP.
\label{twist cP}
\end{equation}
Along the fiber direction,
this indeed acts as
 the permutation  $(X_1, X_2) \, \mapsto \, (X_2, X_1)$.


Let us evaluate the partition function of this permutation orbifold. 
It is convenient to introduce the new coordinates, 
\begin{equation}
X_{\pm} := \frac{1}{\sqrt{2}} \left(X_1 \pm X_2 \right).
\end{equation}
Then, the permutation $\cP$ acts as 
\begin{equation}
\cP ~: ~ X_{\pm} \, \longmapsto \, \pm X_{\pm}. 
\end{equation}
Namely, $\cP$ acts as the ordinary reflection orbifold with respect to the $X_-$-direction.
The insertion of $\cP$ into the trace only leaves the Fock vacua with 
\begin{equation}
p_{-, \, L} \equiv \frac{1}{\sqrt{2}} \left(p_{1,\, L} - p_{2, \, L} \right) = 0, \hspace{1cm} 
p_{-, \, R} \equiv \frac{1}{\sqrt{2}} \left(p_{1,\, R} - p_{2, \, R} \right) = 0 .
\label{cond p-}
\end{equation}
On the other hand, the $X_+$-direction is compactified onto the circle with the radius $\sqrt{2} R_2$,
but the zero-mode spectrum is constrained as
\begin{equation}
\mbox{KK momentum} 
\in 2\bz, \hspace{1cm} \mbox{winding} \in \bz,
\end{equation}
due to the condition \eqn{cond p-}. 
Thus, we eventually obtain the fiber partition function for the sector $\la \equiv m \in 2\bz+1$,
\begin{eqnarray}
\Zf_{\la} (\tau) & = & \left|\frac{2\eta(\tau)}{\th_2(\tau)} \right|
\frac{1}{|\eta(\tau)|^2}\, \sum_{r,s\in \bz}\,
q^{\frac{1}{4} \left(\frac{2r}{\sqrt{2} R_2} + \sqrt{2} R_2 s\right)^2}
\overline{ q^{\frac{1}{4} \left(\frac{2r}{\sqrt{2} R_2} - \sqrt{2} R_2 s\right)^2}}
\nn
& \equiv & \left|\frac{2\eta(\tau)}{\th_2(\tau)} \right| 
\, \sum_{\nu \in \bz \tau + \frac{1}{2} \bz}\, Z_{\sqrt{2} R_2} (\tau\, | \, \nu).
\end{eqnarray}
We also obtain 
\begin{equation}
\Zf_{\la} (\tau) = Z_{R_2}(\tau)^2,
\end{equation}
for $\la = m \in 2\bz$, since $\cP^2 = {\bf 1}$.

Based on these results  and the modular invariance, 
we can uniquely determine the fiber partition functions
$\Zf_{\la} (\tau)$ with $\any \la  \in \La$, which are equal to those of the $I^{(-)}$-model with 
$R_1 = 1/R_2$ and $N=M=1$, as is anticipated. 
In this way, we have seen that the simplest case $N=M=1$ reduces to a geometrically realized model, that is, 
the orbifold by \eqn{twist cP}. 
A similar construction of the partition function involving the permutation 
is given in \cite{Flournoy:2005xe}.
However, in generic cases $NM>1$, our models would be still non-geometric even 
in the $I^{(+)}$-picture, 
because the gluing condition \eqn{gluing G+} 
non-trivially restricts both of the KK and the winding spectra.

~


\section{Unitarized Models}

Let us discuss the unitarity of the  modular invariant 
models we constructed in the previous section. 
The $N=M=1$ model obviously leads to a unitary spectrum as in ordinary $\bz_2$-orbifolds.  
However, in the cases of $NM>1$, the spectrum gets non-unitary. 
For example, one can observe a non-unitary $q$-expansion 
in the sector with base winding $w=N$,%
\footnote
   {Since we are considering the twisted compactification due to  \eqn{def twisting}, 
which includes the shift operator $\cT_{2\pi \Rb}$, we should regard the sectors 
with $w\neq 0$ as the `twisted sectors'
in the context of orbifold theory, whereas  the temporal winding $m$ has 
to be dualized into the KK momentum 
in order to read off the spectrum. 
}
that is, 
\begin{equation}
\la = N  \tau + m, ~~~ (m\in \bz),
\end{equation}
except for the $N=1$ case. 
In fact, we  find  
$S^{(N)}[N\tau] = \bz_N \tau$, 
while $S^{(N)}[N \tau +m] = \bz_N$ holds if $m\neq 0$ and $\gcd \{ |m|, N\} =1$.
Thus, it is not difficult to see that, in this situation,  
we never achieve the $q$-expansion with positive coefficients for the $w=N$ sector, after
making the Poisson resummation with respect to $m\in \bz$.

~


\subsection{Definition of the Unitarized Model}

Therefore, we shall propose the `unitarized' model for the $NM>1$  cases.
Though the complete solution to the unitarization is an important future
problem, one can find a simple and interesting solution which is interpreted
in terms of the {\em discrete torsion\/} \cite{dtorsion} and 
{\em multiple\/} insertions of the interface operators.
Indeed, it turns out that 
one can construct a simple unitary extension of the present model at least in the cases, 
\begin{equation}
R_1= R_2 = \sqrt{\frac{N}{M}} = : \Rf,
\hspace{1cm} \gcd\,\{ N, M \}=1,
\label{R_i MN}
\end{equation}
which satisfy the condition \eqn{cond I-}.
We would like to discuss more general cases with generic $R_1$ and $R_2$ satisfying 
\eqn{cond I-} in the future work.


Let us proceed with the assumption \eqn{R_i MN}.
Our construction of the unitarized model is simple.
All we have to do is to replace the set $S^{(K)}[\la]$ defined in \eqn{SK la} with 
\begin{equation}
\wS^{(K)}[\la] := \{ \nu \in \La\, ;\, \langle \la, \nu \rangle \in K \bz\}/K \La,
\label{hat SK la}
\end{equation}
 in \eqn{Z(0)}, \eqn{Zfeven} and \eqn{rel Zf la bD}.
Namely,  we replace \eqn{Z(0)}  with  
\begin{eqnarray}
\hZ^{(0)}_{\la}(\tau) & := & \sum_{\al \in \wS^{(N)}[\la]}\,  \sum_{\tal \in \wS^{(M)}[\la] } \,
\tau_{\al}^{(N)} \, \ttau_{\tal}^{(M)} \cdot 
Z_{\Rf} (\tau ),
\label{hZ(0)}
\end{eqnarray}
and  the fiber partition functions  $\Zf_{\la}(\tau)$  
with the following:%
\footnote{If we make the replacement \eqn{hat SK la} in \eqn{N M Zf odd}, instead of 
\eqn{rel Zf la bD}, it is subtle if the resultant expression in general 
becomes equal to \eqn{N M Zf odd u}: When  $\gcd \{2N, M\} =2$,  
$\tau^{(2N)}_{\al} \ttau^{(M)}_{\tal}$ may not necessarily be replaced 
with $\tau^{(2N)}_{M \al} \ttau^{(M)}_{2N \tal}$ for 
$\al \in \wS^{(2N)}[\la]$, $\tal \in \wS^{(M)}[\la]$,  and hence not commute.
Even in that case, the expression 
\eqn{N M Zf odd u} is well-defined. 
}
\begin{description}
\item[\underline{even sectors :  $\la  \in 2 \La$}]
\begin{equation}
\hZf_{\la}(\tau) = \hZ^{(0)}_{\la}(\tau)^2.
\label{N M Zf even u}
\end{equation}

\item[\underline{odd sectors : $\la \in \La-2\La$}]
\begin{equation}
\hZf_{\la}(\tau) = \bD_{[\la]} \cdot \left[ \hZ^{(0)}_{2 \la}(\tau)^2\right].
\label{N M Zf odd u}
\end{equation}

\end{description}
The  $\la = 0$ case has been combined into \eqn{N M Zf even u}, since 
$\wS^{(N)}[0] = \La/N\La$, $\wS^{(M)}[0] = \La/M\La$ and thus
\begin{equation}
\hZf_{0}(\tau) = (NM)^2 \, Z_{\frac{M}{N} \Rf}(\tau)^2 \equiv (NM)^2 \,
 Z_{1/\Rf}(\tau)^2
\equiv (NM)^2 \, Z_{\Rf}(\tau)^2.
\end{equation} 
Consequently, the total partition function of the unitarized model is defined by %
%
\begin{eqnarray}
\hspace{-5mm}
\hZ(\tau) 
& : = &  
\sum_{\la \in 2\La}\, \Zb_{\la}(\tau) \, \hZ^{(0)}_{\la}(\tau)^2 
+ \sum_{\la \in \La- 2\La} \, \Zb_{\la}(\tau) \, 
\bD_{[\la]} \cdot \left[ \hZ^{(0)}_{2\la}(\tau) \right]^2.
\label{hZtot N M}
\end{eqnarray}
Again one can easily check that this function reduces to the total partition function given in section \ref{N=M=1},
when setting $N=M=1$.

From the definition, it follows that $S^{(K)}[\la] \subset \wS^{(K)}[\la]$ in general,
and that 
\begin{eqnarray}
&& S^{(K)}[\la] = \wS^{(K)}[\la], ~~~ \mbox{iff} ~~ 
\gcd \{|w|, |m|, K\} =1, 
~ (\la \equiv w \tau +m \in \La), ~~ \mbox{or} ~ \la=0.
\label{hat SK rel}
\end{eqnarray}
In other words, new twisted sectors are created when $ \gcd \{|w|, |m|, K\} >1$.
We note that $\wS^{(K)}[\la]$ has a periodicity, 
\begin{equation}
\wS^{(K)}[\la + K \la' ] = \wS^{(K)}[\la ], \hspace{1cm} (\any \la' \in \La).
\end{equation}

As mentioned above and discussed below,
the replacement by  $\wS^{(K)}[\la]$ 
is  interpretable  in terms of the  discrete torsion
in the  theory of orbifolds, which is expected to preserve the unitarity. 
This fact also motivated us to define the unitarized model in such a  way. 
%
Furthermore, we will discuss later how the new partition function
$\hZf_{\la}(\tau)$ is interpreted to be generated by  multiple insertions of 
$\cI^{(-)}$  along various cycles of the world-sheet torus.


~

\subsection{Proof of Unitarity}

We here demonstrate that the proposed model given by \eqn{hZtot N M}
is unitary. Namely, we show  that the total partition function \eqn{hZtot N M} 
is indeed $q$-expanded 
only with positive integral coefficients,  after making the Poisson resummation with respect to 
$m\in \bz$ in $\la \equiv w\tau + m$. 

We proceed by three steps: First, 
 since the total partition 
function looks quite intricate, we analyze a simpler 
part of the partition function in \eqn{hZ(0)}.
Second,  using the result in the first step, we show that 
the sectors with even $w$ in $ \la = w\tau + m$ have
a $q$-expansion with positive integral coefficients.
Finally, we show a similar statement for the sectors with odd $w$.

~


\noindent
\underline{\bf (i) evaluation of $\hZ^{(0)}_{\la}(\tau)$ :}

We first examine the simpler partition function $\hZ^{(0)}_{\la}(\tau)$ defined in \eqn{hZ(0)}.
With this aim it is convenient to introduce the phase factor defined by
\begin{eqnarray}
&& \ep^{(K)}_{j} (\la, \gamma ; \tau ) := e^{- 2\pi i \frac{j}{K} \langle \la , \gamma \rangle_{\tau}}
\equiv e^{- 2\pi i \frac{j}{K} (w b-ma)},
\nn
&& 
\hspace{2cm}
 (j \in \bz_K , ~ \la\equiv w\tau+m \in \La, ~ \gamma \equiv a\tau+b \in \La),
\label{ep j}
\end{eqnarray}
where 
$\langle ~, ~ \rangle_{\tau}$ is defined in \eqn{<>} (see also Appendix A). 
By the subscript, we have made    
the $\tau$-dependence explicit to avoid confusion.
Since 
\begin{equation}
\frac{1}{K} \sum_{j\in \bz_K}\, \ep^{(K)}_{j} (\la, \gamma ; \tau )
= \left\{
\begin{array}{ll}
1 & ~~ \langle \la, \, \gamma \rangle_{\tau} \in K\bz, \\
0 & ~~ \mbox{otherwise},
\end{array}
\right.
\end{equation}
we can rewrite $\hZ^{(0)}_{\la}(\tau)$ as
\begin{eqnarray}
\hZ^{(0)}_{\la}(\tau) &=&  
\sum_{j\in \bz_N}\, \sum_{\tj \in \bz_M}\, 
\hZ^{(0)}_{\la, \, [j,\tj]} (\tau),
\end{eqnarray}
with 
\begin{eqnarray}
\hspace{-1cm}
\hZ^{(0)}_{\la,\, [j, \tj]}(\tau) &: =& \frac{1}{NM} 
\sum_{\al \in \La/N\La} \, \sum_{\tal \in \La/M\La}\,
\ep_j^{(N)} (\la, \al; \tau) \, \ep_{\tj}^{(M)} (\la, \tal; \tau)\, 
\tau_{\al}^{(N)} \ttau_{\tal}^{(M)} \cdot Z_{\Rf}(\tau)
\nn
&\equiv & \frac{1}{NM} 
\sum_{\al \in \La/N\La} \, \sum_{\tal \in \La/M\La}\,
\ep_j^{(N)} (\la, M \al ; \tau) \, \ep_{\tj}^{(M)} (\la, N \tal; \tau)\, 
\tau_{M \al}^{(N)} \ttau_{N \tal}^{(M)} \cdot Z_{\Rf}(\tau).
\label{def hZj}
\end{eqnarray}
In the second line, we made use of the assumption $\gcd \{ N, M\} = 1$.
Note here that $\tau^{(N)}_{M \al}$ and $\ttau^{(M)}_{N \tal}$ always commute with 
each other, while $\tau^{(N)}_{\al}$ and $\ttau^{(M)}_{\tal}$
do not.

The phase factor
$\ep^{(N)}_{j} (\la, \gamma ; \tau) \ep^{(M)}_{\tj} (\la, \gamma ;\tau)$ is interpreted 
as the {\em discrete torsion\/} \cite{dtorsion}, as we addressed before.
Actually, it turns out that the sector of the partition function \eqn{hZtot N M}
with $w$ even or odd  separately leads  to a unitary $q$-expansion.
To see this below, it is important to observe that  \eqn{def hZj} is explicitly evaluated as
\begin{align}
 \hZ^{(0)}_{\la= w\tau+m,  \, [j,\tj]} (\tau)  =  
\sum_{a \in \bz_N} \, \sum_{\ta \in \bz_M}\, \sum_{r \in \bz_2}
e^{ 2\pi i m \left( \frac{j M a}{N} + \frac{\tj N \ta}{M}  \right)} \,
Z^{(NM)}[M j w + N \ta + NMr, \, N \tj w + M a ] (\tau).
\label{eval hZj}
\end{align}
Here, we set $\al = a\tau + b$, $\tal = \ta \tau + \tb$, summed over $b \in \bz_N$, $\tb \in \bz_M$ 
and used the notation \eqn{Zuv}. 
This expression is suited for making the Poisson resummation, 
since the temporal winding $m$ appears 
only in the phase factor $e^{2\pi i m \left( \frac{j M a}{N} + \frac{\tj N \ta}{M}  \right)}$.

~


\noindent
\underline{\bf (ii) unitarity of $w\in 2\bz$ sectors :}

Now, we are ready to show the unitarity of the model or the partition function in question, that is, 
 $\hZ(\tau)$  in \eqn{hZtot N M}. 
We first focus on the case of $\la = w\tau+m$ with a fixed value $w\in 2\bz$,
and consider the Poisson resummation with respect to $m$.

For the cases of $m \in 2\bz$, the fiber partition function is equal to the square of 
\eqn{hZ(0)}, that is, 
\begin{eqnarray}
 \hZf_{\la}(\tau) & = & \prod_{i=1,2}\, \sum_{j_i \in \bz_N}\, \sum_{\tj_i \in \bz_M} \, 
\hZ_{\la, \, [j_i, \tj_i]}(\tau).
\label{eval hZf 1}
\end{eqnarray}
Then, it is straightforward to make the Poisson resummation over $m \in 2\bz$ 
with the help of \eqn{eval hZj}.
Note that the phase factor appearing in \eqn{eval hZj} 
just shifts the KK momentum along the  base circle as 
\begin{equation}
\frac{n}{2 \Rb} ~ \longrightarrow ~ 
\frac{1}{2 \Rb} \left[ n +
2 \sum_{i=1,2}\, \left(
\frac{j_i M a_i}{N} + \frac{\tj_i N \ta_i }{M}\right) \right],
\end{equation}
where $i=1,2$ labels the contributions from the two factors of $\hZ^{(0)}_{\la}(\tau)$.
We thus obtain
\begin{eqnarray}
&& 
\hspace{-1cm}
\hZ_{\msc{even}, \, w}(\tau) \equiv \sum_{m \in 2\bz}\, \Zb_{w\tau+m}(\tau)\, \hZf_{w\tau+m}(\tau)
\nn
&& 
\hspace{5mm}
= 
 \sum_{j_i,a_i \in \bz_N} \, \sum_{\tj_i,\ta_i \in \bz_M} \, \sum_{r_i \in \bz_2}\,  \sum_{n\in \bz}\,
\left[ \prod_{i=1,2} \,
Z^{(NM)}[M j_i w + N \ta_i + NMr_i, \, N \tj_i w + M a_i ] (\tau) \right]
\nn
&& 
\hspace{5mm}
\times \frac{1}{2} \, 
\frac{1}{\left|\eta(\tau)\right|^2}\,
q^{\frac{1}{4} \left[\frac{1}{2\Rb}
\left\{ n +
2 \sum_{i=1,2}\, \left(
\frac{j_i M a_i}{N} + \frac{\tj_i N \ta_i }{M}\right)\right\} +  \Rb w
\right]^2}\,
\overline{
q^{\frac{1}{4} \left[\frac{1}{2\Rb}
\left\{ n +
2 \sum_{i=1,2}\, \left(
\frac{j_i M a_i}{N} + \frac{\tj_i N \ta_i }{M}\right)\right\} -  \Rb w
\right]^2}
}.
\nn
&&
\label{PR hZ even w}
\end{eqnarray}
It is not difficult to see that the sectors with `non-diagonal' quantum numbers 
$(j_1, a_1 , \ldots) \neq (j_2, a_2, \ldots)$
are expanded in a $q$-series with positive integral coefficients. Note  that 
the factor $\frac{1}{2}$ appearing in \eqn{PR hZ even w} indeed cancels out due to 
the degeneracy of relevant contributions.
Furthermore,  the partition functions 
for the `diagonal' sectors with  
$j_1 = j_2 \equiv j$, $a_1 = a_2 \equiv a$, and so on, 
are rewritten by using the product formula of theta function \eqn{product} as 
\begin{eqnarray}
&& 
\hZ_{\msc{even}, \, w}[j, \tj, a, \ta, r](\tau) = 
\frac{1}{\left|\eta(\tau)\right|^2} \,Z^{(2NM)}[2 M j w + 2 N \ta + 2 NMr, \, 2 N \tj w +2 M a ] (\tau)
\nn
&&
\hspace{2cm}
\times \frac{1}{2} \,
\sum_{n\in \bz} \,
\frac{1}{\left|\eta(\tau)\right|^2}\,
q^{\frac{1}{4} \left[\frac{1}{2\Rb}
\left\{ n +
4 \left(
\frac{j M a}{N} + \frac{\tj N \ta }{M}\right)\right\} +  \Rb w
\right]^2}\,
\overline{
q^{\frac{1}{4} \left[\frac{1}{2\Rb}
\left\{ n +
4 \, \left(
\frac{j M a}{N} + \frac{\tj N \ta }{M}\right)\right\} -  \Rb w
\right]^2}
}
\nn
&&
\hspace{2cm}
+ \ [\mbox{residual terms}],
\label{hZ even w diag 1}
\end{eqnarray}
The `residual terms' in the above denote some  $q$-series with positive integral coefficients
 whose explicit form is not important here.

On the other hand,  by using the definition of 
$\bD_{[\la]}$ given in \eqn{def bDwm 2},
we obtain for $m \in 2\bz+1$, 
\begin{eqnarray}
\hZf_{\la}(\tau) & = & \bD_{[\la]} \cdot \left[
\prod_{i=1,2}\, \sum_{j_i \in \bz_N}\, \sum_{\tj_i \in \bz_M} \, 
\hZ_{2 \la, \, [j_i, \tj_i]}(\tau) \right]
\nn
& = & \sum_{j\in \bz_N} \, \sum_{\tj \in \bz_M} \, \hZ^{(0)}_{2 \la, \, [j,\tj]}(2\tau)
\nn
& = & \sum_{j\in \bz_N} \, \sum_{\tj \in \bz_M} \,
\sum_{a \in \bz_N} \, \sum_{\ta \in \bz_M}\, \sum_{r \in \bz_2}
e^{ 2 \pi i m \cdot 2 \left( \frac{j M a}{N} + \frac{\tj N \ta}{M}  \right)} 
\nn
&& 
\hspace{1cm}
\times 
Z^{(NM)}[M j w + N \ta + NMr, \, N \tj w + M a ] (2 \tau)
\nn
&= & \sum_{j\in \bz_N} \, \sum_{\tj \in \bz_M} \,
\sum_{a \in \bz_N} \, \sum_{\ta \in \bz_M}\, \sum_{r \in \bz_2}
e^{ 2 \pi i m \cdot 2 \left( \frac{j M a}{N} + \frac{\tj N \ta}{M}  \right)} 
\nn
&& 
\hspace{1cm}
\times 
\left| \frac{2\eta(\tau)}{\th_2(\tau)} \right|
\,Z^{(2 NM)}[2 M j w + 2 N \ta + 2 NMr, \, 2N \tj w + 2 M a ] (\tau).
\label{eval hZf 2}
\end{eqnarray}
Note here that the operator $\bD_{[\la]}$ for this sector acts 
as $\tau\, \rightarrow \, 2\tau$, {\em while 
keeping $\la \equiv w\tau+m$ intact.} 
This implies that 
\begin{eqnarray}
&& \ep_j^{(N)} (2 \la, M \left\{ a (2\tau) +b\right\}; 2\tau) \, \ep_{\tj}^{(M)} 
(2 \la, N \{ \ta (2 \tau) +\tb\} ; 2\tau) 
\nn
&& \hspace{3cm}
 =  
e^{-2\pi i \frac{j M}{N} \langle w (2\tau)+2 m, \, a (2\tau) +b \rangle_{2 \tau}} \, 
e^{-2\pi i \frac{\tj N}{M} \langle w (2\tau)+ 2m, \, \ta (2\tau) +\tb \rangle_{2 \tau}}
\nn
&& \hspace{3cm} 
=  e^{-2\pi i \frac{j M}{N} \left(w b- 2 ma\right)} \, e^{-2\pi i \frac{\tj N}{M} \left(w \tb- 2 m\ta\right)},
\end{eqnarray} 
which explains the phase factor appearing in \eqn{eval hZf 2}.
As a consistency check, 
we can confirm the modular T-invariance of \eqn{eval hZf 2}, 
\begin{equation}
\hZf_{\la}(\tau+1) = \hZf_{\la}(\tau),
\hspace{1cm} (\any \la \equiv w\tau+m, ~ w \in 2\bz, ~ m \in 2\bz+1).
\end{equation}


The expressions \eqn{eval hZf 2} is again suitable for the Poisson resummation. 
Namely, the summation over $m \in 2\bz+1$ is evaluated as 
\begin{eqnarray}
\hZ_{\msc{odd}, \, w} (\tau) &\equiv& 
\sum_{m \in 2\bz+1} \, \hZf_{w\tau +m} (\tau) \, \Zb_{w\tau+m}(\tau)
\nn
& \equiv & \sum_{j,a \in \bz_N} \, \sum_{\tj, \ta \in \bz_M}\, \sum_{r\in \bz_2}\,
\hZ_{\msc{odd}, w} [j, \tj, a, \ta, r](\tau) ,
\end{eqnarray}
with
\begin{eqnarray}
&& 
\hspace{-1cm}
\hZ_{\msc{odd}, \, w} [j, \tj, a, \ta, r](\tau) 
= \left| \frac{2\eta(\tau)}{\th_2(\tau)} \right|\, 
Z^{(2 NM)}[2M j w + 2 N \ta + 2NMr, \, 2 N \tj w + 2 M a ] (\tau)
\nn
&& \hspace{5mm}
\times  \sum_{n\in \bz} \, \frac{1}{2}
\frac{(-1)^n}{\left|\eta(\tau)\right|^2}\,  
q^{\frac{1}{4} \left[\frac{1}{2\Rb}
\left\{ n +
 4 \left( \frac{j M a}{N} + \frac{\tj N \ta }{M}\right)\right\} +  \Rb w
\right]^2}\,
\overline{
q^{\frac{1}{4} \left[\frac{1}{2\Rb}
\left\{ n +
 4\left(\frac{j M a}{N} + \frac{\tj N \ta }{M}\right)\right\} -  \Rb w
\right]^2}
}.
\label{PR hZ w odd}
\end{eqnarray}

Comparing \eqn{hZ even w diag 1} and \eqn{PR hZ w odd}, 
it is now obvious that the summation 
\begin{equation}
\hZ_{\msc{even}, \, w} [j, \tj, a, \ta, r](\tau) + \hZ_{\msc{odd}, \, w} [j, \tj, a, \ta, r](\tau) ,
\label{hZeven+hZodd w}
\end{equation}
is written in a unitary $q$-series for each $j$, $\tj$, $a$, $\ta$, $r$. 

~

\noindent
\underline{\bf (iii) unitarity of $w\in 2\bz+1$ sectors :}

Let us examine the remaining cases with $w\in 2\bz+1$. 
By means of \eqn{def bDwm 2}, we obtain the following:
\begin{itemize}
\item For $m \in 2\bz$,
\begin{eqnarray}
\hspace{-5mm}
\hZf_{\la}(\tau) & = & \bD_{[\la]} \cdot \left[
\prod_{i=1,2}\, \sum_{j_i \in \bz_N}\, \sum_{\tj_i \in \bz_M} \, 
\hZ_{2 \la, \, [j_i, \tj_i]}(\tau) \right]
\nn
& = & \sum_{j\in \bz_N} \, \sum_{\tj \in \bz_M} \, \hZ^{(0)}_{ \frac{1}{2}(2\la) , \, [j,\tj]}\left(\frac{\tau}{2}\right)
\nn
& = & \sum_{j\in \bz_N} \, \sum_{\tj \in \bz_M} \,
\sum_{a \in \bz_N} \, \sum_{\ta \in \bz_M}\, \sum_{r \in \bz_2}
e^{ 2 \pi i m \left( \frac{j M a}{N} + \frac{\tj N \ta}{M}  \right)} 
\nn
&& 
\hspace{1cm}
\times 
Z^{(NM)}\left[2 M j w + N \ta + NMr, \, 2 N \tj w + M a \right] \left( \frac{\tau}{2} \right).
\label{eval hZf 3}
\end{eqnarray}
\item For $m \in 2\bz+1 $,
\begin{eqnarray}
\hspace{-5mm}
\hZf_{\la}(\tau) & = & \bD_{[\la]} \cdot \left[
\prod_{i=1,2}\, \sum_{j_i \in \bz_N}\, \sum_{\tj_i \in \bz_M} \, 
\hZ_{2 \la, \, [j_i, \tj_i]}(\tau) \right]
\nn
& = & \sum_{j\in \bz_N} \, \sum_{\tj \in \bz_M} \, 
\hZ^{(0)}_{ \frac{1}{2}(2\la), \, [j,\tj]}\left(\frac{\tau+1}{2}\right)
\nn
& = & \sum_{j\in \bz_N} \, \sum_{\tj \in \bz_M} \,
\sum_{a \in \bz_N} \, \sum_{\ta \in \bz_M}\, \sum_{r \in \bz_2}
e^{ 2 \pi i (m-w)  \left( \frac{j M a}{N} + \frac{\tj N \ta}{M}  \right)} 
\nn
&& 
\hspace{1cm}
\times 
Z^{(NM)}\left[2 M j w + N \ta + NMr, \, 2 N \tj w + M a \right] \left( \frac{\tau+1 }{2} \right).
\label{eval hZf 4}
\end{eqnarray}
\end{itemize}
In these evaluations, the phase factors were slightly non-trivial again. 
In \eqn{eval hZf 3}, for example,
we made the following evaluation, 
\begin{eqnarray}
&& \ep_j^{(N)} \left(\la, M \{ a \frac{\tau}{2} +b\} ; \frac{\tau}{2} \right) \, 
\ep_{\tj}^{(M)} \left(\la, N \{ \ta \frac{\tau}{2} +\tb\}; \frac{\tau}{2} \right)
\nn
&& \hspace{4cm} =  
e^{-2\pi i \frac{j M}{N} \langle 2w\frac{\tau}{2} +m, \, a \frac{\tau}{2} +b \rangle_{\frac{\tau}{2}}} \, 
e^{-2\pi i \frac{\tj N}{M} \langle 2 w\frac{\tau}{2}+m, \, \ta \frac{\tau}{2}+\tb \rangle_{\frac{\tau}{2}}}
\nn 
&& \hspace{4cm}
= 
e^{-2\pi i \frac{j M}{N} \left(2wb- m a\right)} \, 
e^{-2\pi i \frac{\tj N}{M} \left(2w\tb- m\ta\right)}.
\end{eqnarray}

If only picking \eqn{eval hZf 3} up, 
we are clearly led to a unitary $q$-series 
by Poisson resumming over $m\in 2\bz$, as in the case of 
$\hZ_{\msc{even}, \, w}(\tau)$ in \eqn{PR hZ even w}.
 Moreover, as expected, 
\eqn{eval hZf 3} and \eqn{eval hZf 4} are combined into a T-invariant form, 
since
\begin{equation}
\hZf_{\la}(\tau+1) = \hZf_{\la} (\tau), 
\hspace{1cm} (\la \equiv w\tau+m, ~ \any w\in 2\bz+1, ~ \any m \in \bz).
\end{equation}
This means that the summation over $m \in 2\bz+1$
is written in the same form of $q$-series 
as that for $m \in 2\bz$, but {\em with an extra sign $\pm 1$ in each term\/}.
Therefore, the partition function 
\begin{eqnarray}
&& \sum_{m\in  \bz}\, \hZf_{w\tau+m} (\tau) \, \Zb_{w\tau+m} (\tau) 
\nn
& &
\hspace{1cm}
\equiv
\sum_{m\in 2\bz}\,\left[ \hZf_{w\tau+m} (\tau) \, \Zb_{w\tau+m} (\tau) + 
 \hZf_{w(\tau+1)+m} (\tau+1) \, \Zb_{w(\tau+1)+m} (\tau+1) \right], \quad 
\end{eqnarray}
again produces a  unitary $q$-series.
Similarly to the twisted sectors of the familiar $\bz_2$-orbifolds,
the negative terms due to the oscillator part cancel with the 
corresponding positive terms. 

In this way, we have succeeded  in showing the unitarity of the proposed model \eqn{hZtot N M}.


~

We add a few comments:
\begin{itemize}
\item
It is worthwhile to point out that there exists 
a {\em unique\/}  vacuum with minimal conformal weights 
$h=\tilde{h}=0$ (the `identity state')
in the spectrum read off from the partition function \eqn{hZeven+hZodd w}
for the sector $w=j=\tj=a=\ta = r=0$. This fact would suggest that our choice 
of the normalization constants 
\eqn{cN la} is reasonable.


\item
As clarified in the above discussion, 
the even sectors with $\la \in 2 \La$ are reinterpreted as the orbifold with 
the discrete torsion. 
Nevertheless, the total partition function \eqn{hZtot N M} is not 
likely to be described by any model of orbifold in the precise sense, 
except for the simplest case $N=M=1$. 
This is because the operator  $\bD_{[\la]}$ is hard to be explicitly 
realized in terms of  an automorphism acting on the Hilbert space of CFT. 

\end{itemize}

~


\subsection{Interpretation of the Unitarized Model : Multiple Insertions of the Interface Operators}
\label{multiple}

At the last of this section we would like to discuss a possible physical interpretation 
of the unitarized model described by  \eqn{hZtot N M},
or the fiber partition functions $\hZf_{\la}(\tau)$ given in \eqn{N M Zf even u} 
and \eqn{N M Zf odd u}. 
Let us start by extending the formulas \eqn{Zf cI even}, \eqn{Zf cI odd} to the cases of 
multiple insertions of the interface operator $\cI^{(-)}$ along different cycles, 
\begin{eqnarray}
&&
\hspace{-1.5cm}
\Zf[\cI^{(-)}_{\gamma_1}, \ldots, \cI^{(-)}_{\gamma_\ell}](\tau) :=  
\left[ \left( \prod_{j=1}^{\ell} \, \bP^{(N)}_{\gamma_j} \btP^{(M)}_{\gamma_j} \right) \cdot 
Z_{\sqrt{\frac{N}{M}}} (\tau) \right]^2,
~~~ (\gamma_1, \ldots, \gamma_{\ell} \in 2\La-\{0\}),
\label{Zf cI multiple even}
\\
&&
\hspace{-1.5cm}
\Zf[\cI^{(-)}_{\gamma}, \cI^{(-)}_{\gamma_1}, \ldots, \cI^{(-)}_{\gamma_\ell}](\tau) := 
\bD_{[\gamma]} \cdot
\left[ \bP^{(N)}_{2\gamma} \btP^{(M)}_{2\gamma} 
\left( \prod_{j=1}^{\ell} \, \bP^{(N)}_{\gamma_j} \btP^{(M)}_{\gamma_j} \right) \cdot Z_{\sqrt{\frac{N}{M}}} (\tau) \right]^2,
\nn
&&
\hspace{7cm}
 (\gamma \in \La -2 \La, ~~\gamma_1, \ldots, \gamma_{\ell} \in 2\La-\{0\}).
\label{Zf cI multiple odd even}
\end{eqnarray}
These expressions are  well-defined because the operators $\bP^{(N)}_*$, $\btP^{(M)}_*$ commute with one another, 
as we already mentioned. 
One should keep it in mind that all the operators $\tau^{(N)}_*$, $\ttau^{(M)}_*$ implicitly 
appearing in \eqn{Zf cI multiple even}, \eqn{Zf cI multiple odd even}  have to be interpreted as   $\tau^{(N), \Rf}_*$, 
$\ttau^{(M), \Rf}_*$ respectively. 
At present, we do not have a 
proper extension for 
any two insertions along different {\em odd\/}  cycles;
$
\gamma_i, \gamma_j \in \La- 2\La, ~
 \gamma_i \neq \gamma_j.
$

Now, let us focus on the partition functions 
$\hZf_{\la}(\tau)$ given in \eqn{N M Zf even u}, \eqn{N M Zf odd u}. 
We attempt to rewrite them in the forms of \eqn{Zf cI multiple even} 
and \eqn{Zf cI multiple odd even}.
Recall that the partition function in the even sector \eqn{N M Zf even u} is 
obtained by replacing the sets $S^{(*)}[\la]$ appearing in \eqn{N M Zf even}
with their `hatted' counterparts \eqn{hat SK la}, and 
in the odd sector by acting with the diagonal operator $\bD_{[\la]}$.

We again begin our analysis  with 
the decomposition $\la = n_{\la} \la_0$ for $\any \la \in \La- \{0\}$ with $ \la_0 \in \La_0$, $n_{\la} \in \bz$.
We also introduce the unique element 
$\la_1 \in \La_0$ 
by requiring the condition $\langle \la_1, \la_0 \rangle =1$. 
Then, making use of the assumption $\gcd \{ N , \, M\} =1$, we find  
\begin{eqnarray}
\wS^{(K)}[\la] & = & \bz_K \la_0 + \bz_{n_{\la}^{(K)}} \frac{K}{n^{(K)}_{\la}} \la_1
\nn
& = & S^{(K)}[\la] + S^{(K)}[\la+ NM \la_1] ,
\hspace{1cm} 
(K=N, ~  M),
\label{rel wS S}
\end{eqnarray}
where we set
\begin{equation}
n_{\gamma}^{(K)} := \gcd \{ |n_{\gamma}|, K \}, \hspace{1cm} (K= N, M).
\label{def n gamma}
\end{equation}
For instance, in the case of $\la = m \in \bz -\{0\}$, it is easy to see
\begin{equation}
\wS^{(N)}[m]  =  \bz_N  + \bz_{n_{m}^{(N)}}  \frac{N}{n_{m}^{(N)}} \tau
=  S^{(N)}[m] + S^{(N)}[m + NM \tau],
\end{equation}
and the equality for  general case $\any \la \in \La-\{0\}$
follows from modular transformations.  

From the above relations, 
we obtain the following identity, 
\begin{equation}
\left( n_{\la}^{(N)} N\right)  \, \bP^{(N)}_{\la} \bP^{(N)}_{\la + NM \la_1}  
= \sum_{\al\in \wS^{(N)}[\la]}\,\tau^{(N)}_{\al},
\label{rel bP bhP}
\end{equation}
and a similar relation for $\btP^{(M)}_*$.
The equality \eqn{rel bP bhP} implies that 
$\hZf_{\la}(\tau)$  given in \eqn{N M Zf even u}  for the even sector $\la \in 2\La -\{0\}$  is 
identified with the partition function of the type \eqn{Zf cI multiple even} with 
the insertions of $\cI^{(-)}_{\la}$ and $\cI^{(-)}_{2(\la + NM \la_1)}$ 
up to a normalization constant. 
To be more precise, it is straightforward to show 
\begin{eqnarray}
\hZf_{\la}(\tau) &=& \cN_{\la}' \, \Zf[\cI^{(-)}_{\la}, \cI^{(-)}_{2(\la + NM \la_1)}] (\tau)
\nn
& \equiv &
\left( n_{\la}^{(N)} n_{\la}^{(M)} NM\right)^2 \, 
\Zf[\cI^{(-)}_{\la}, \cI^{(-)}_{2(\la + NM \la_1)}] (\tau), \hspace{0.7cm} (\any \la \in 2\La -\{0\}).
\label{hZf cI even}
\end{eqnarray}
The factor $2$ of the subscript in the latter operator assures that
the cycle is even, though the actual action is  implemented also by  
$\bP^{(N)}_{\la + NM \la_1} $ due to \eqn{Pnla}. 
Furthermore, the partition function for the odd sector \eqn{N M Zf odd u}
is rewritten in the form of \eqn{Zf cI multiple odd even}, that is,  
\begin{eqnarray}
\hspace{-5mm}
\hZf_{\la}(\tau) 
& = & \cN'_{\la} \, 
\Zf[\cI^{(-)}_{\la}, \cI^{(-)}_{2(2\la + NM \la_1)}] (\tau)
\nn
& \equiv & 
\bD_{[\la]} \cdot \left[
\cN_{2\la}' \, \Zf[\cI^{(-)}_{2\la}, \cI^{(-)}_{2(2\la + NM \la_1)}] (\tau)
\right]
\nn
& \equiv & 
\left( n_{2\la}^{(N)} n_{2\la}^{(M)} NM \right) \, 
\bD_{[\la]}\cdot 
\Zf[\cI^{(-)}_{2 \la}, \cI^{(-)}_{2(2\la + NM  \la_1)}] (\tau),
\hspace{0.7cm} (\any \la \in \La - 2\La).
\label{hZf cI odd}
\end{eqnarray}
These are the equalities we have looked for,  
and show that the fiber partition functions $\hZf_{\la}(\tau)$ are interpreted 
to be generated by multiple insertions of the interface operators along different cycles.

~


\section{Summary and Discussion}

In this paper,  we studied simple models of non-geometric backgrounds 
based on the world-sheet CFT equipped with the topological interfaces
\cite{Petkova:2000ip,Bachas:2004sy}.
The topological interfaces there composed
the twist operators similar to those in orbifold CFTs, 
and we performed the twisted compactification of the Scherk-Schwarz type \cite{SS1,SS2}. 
When crossing the defect lines on the world-sheet  
created by the interface operators, the KK momenta and the winding modes
are mixed, which embodies a `non-geometry' (or the `stringy geometry', possibly).
Requiring the modular invariance, we needed to sum up all the 
winding sectors with respect to the Scherk-Schwarz circle, 
 or the base space.  The fiber CFT, on the other hand,
was  made up of the world-sheet with the interfaces aligned suitably, 
so as to correlate  in a modular covariant
manner to the winding numbers of the base circle.

In our concrete studies, 
two types of the topological interfaces appeared.
One is the group-like defect and the other, which includes the former,  is 
the duality defect 
\cite{Frohlich:2004ef,Frohlich:2006ch}.
The group-like defect corresponds to the case of $N=M=1$ 
in section \ref{N=M=1},  and 
resulted in  an example of the T-fold, that is, the Scherk-Schwarz 
compactification twisted by the T-duality. 
As we clarified in section \ref{I+}, this model is also T-dualized to  
the permutation orbifold,  which  manifestly yields a unitary theory.

On the other hand, the cases of $NM>1$ given in section \ref{NM>1}
are described in terms of the duality defects that 
do not implement symmetries in the strict sense. 
The obtained partition function \eqn{Ztot N M}
looks rather intricate and the unitarity is not realized as it stands.
As mentioned in section \ref{TopInt}, 
the topological interfaces in these cases 
implement a projection that restricts the zero-mode spectrum, instead of 
an automorphism of the Hilbert space.
In order to maintain the unitarity, or conserve the probability, one may thus need 
to complement this projection by supplying `twisted sectors' which may differ 
from those in ordinary orbifold theories.
Quite interestingly,  a `unitarization' was indeed possible at least for 
the cases of $R_1=R_2 = \sqrt{\frac{N}{M}}$, 
as demonstrated in section 4. Furthermore, the unitarized model is 
found to be described 
by the world-sheet with multiple insertions of the interfaces along different cycles. 
While the even sectors of this model might 
be identified with some orbifold with the discrete torsion 
\cite{dtorsion},  the total partition function itself does not seem to correspond to any 
orbifold conformal theory.%
\footnote{
It would be worth mentioning that, in the papers 
\cite{Frohlich:2009gb,Carqueville:2012dk,Brunner:2013ota},
the authors have been investigating  the `generalized orbifolds' that do not arise from any symmetry group, for example,   
in the context of the  topological Landau-Ginzburg theories with defects.
}

We organized the interface operators, as mentioned in section 
\ref{section:ConstructionModel}, 
so that they act 
consistently within one Hilbert space. This is in accord with the low-energy analysis 
that the T-folds lie at the fixed points of the moduli space. Taking also into account 
our motivation to discuss possible roles of the world-sheet interfaces for string theory, 
our models may be the simplest from our point of view. Compared with the preceding 
works \cite{Flournoy:2005xe,Hellerman:2006tx,Kawai:2007qd,Condeescu:2012sp,
Condeescu:2013yma}, where the exact CFT partition functions for T-folds are 
constructed by identifying some asymmetric twists with particular T-duality 
transformations, our construction takes a different route based on the interfaces.
In the case of the group-like defect, our resultant model, however, fits into the same 
category of asymmetric orbifolds, 
except that our model has a continuous modulus in a fixed line instead of 
isolated fixed points. In the case of the duality defects, 
our construction would be a novel type other than ordinary asymmetric orbifolds.
In both cases, our approach would provide a unified picture 
for the non-geometric backgrounds of T-fold type.  
Our approach may also be generalized 
to a large class of models, as long as the action 
of the interfaces is well-defined within one Hilbert space, which implies that the model 
is on the fixed submanifold in the moduli space.

The analysis of non-geometric backgrounds typically involves
compactification radii of or below the string scale. The advantage of
the world-sheet CFT approach is that the results are $\alpha'$-exact and thus
valid even at the string scale, where the notion of classical geometry may not be valid.
When the string coupling for the genus expansion becomes large,
the world-sheet approach here is not applicable,  
and should be superseded by non-perturbative approaches.

For future work,
if intending the applications to string compactifications,  
we would have several tasks to be done: 

\begin{itemize}
\item the construction of the unitarized models for 
general radii $R_1$, $R_2$ satisfying the condition \eqn{assumption R1R2};

\item  the extension to the case of the fiber CFT on higher dimensional tori;

\item the supersymmetric extension by means of superconformal interfaces;

\item the target-space interpretation of the insertions of the interfaces and the 
 resultant models.

\end{itemize}

Especially, the topological interfaces in $\cN=1$ SCFTs on higher dimensional tori 
have been elaborated in \cite{Bachas:2012bj}.
It has been  clarified there that the duality (or the `orbifold equivalence') defects
which do not preserve the charge lattice of string theory 
generically correspond to the  $O(d,d ; \bq)$-group 
for $d$-dimensional torus. It is an 
 interesting question whether one can construct 
the unitary and modular invariant models based on such $O(d,d ; \bq)$-interfaces
as extensions of the unitarized model proposed in this paper. 
It would also be interesting to see, as mentioned in \cite{Bachas:2012bj},  that 
an object representing the `quasi-symmetry' survives 
even under the $\alpha'$ corrections and 
plays a definite role in string theory.

Moreover, a natural direction of the future studies would be 
 the extensions to non-trivial
cases with $\cN=2$ SCFTs, say, the Gepner models as the fiber CFT, 
in which one would handle the duality defects 
inducing the mirror transformations.
Of course, one may imagine  the models of Scherk-Schwarz compactifications  
with the {\em self-dual\/} mirror twisting acting on some $\cN=4$ fiber SCFTs, 
similarly to  \cite{Kawai:2007nb}. These  cases  may 
be described by the group-like defects and reduce to asymmetric orbifolds. 
However,  more general duality defects made up of the $\cN=2$ topological interfaces 
would be capable of producing a much broader class of superstring vacua,  
which would  be fairly non-trivial and curious.

Another interesting direction to be pursued would be 
the possibility of more general unitary models 
by multiple insertions of the interface operators. In section \ref{multiple}, we chose  
particular cycles along  which   the  interfaces are aligned,  in order to 
obtain the unitarized model
\eqn{hZtot N M}. 
However,  it is not yet clear what
the general principle is, in order for 
the configuration of the interfaces to realize sensible unitary models. 
In any case,  we hope that this work would serve as 
a step to construct a novel type of string vacua based on the world-sheets 
equipped with the topological interfaces or the interface operators composed of them.

~


\section*{Acknowledgments}

We would like to thank K. Sakai for useful comments on the conformal interfaces.
This work is supported in part by 
JSPS KAKENHI Grant Number 24540248 and 23540322
from Japan Society for the Promotion of Science (JSPS).


~

\section*{Appendix A: ~ Conventions for Theta Functions and Some Useful Formulas}

\setcounter{equation}{0}
\def\theequation{A.\arabic{equation}}

\noindent
\underline{\bf Theta functions:}
%
 \begin{align}
 & \dsp \th_1(\tau,z)=i\sum_{n=-\infty}^{\infty}(-1)^n q^{(n-1/2)^2/2} y^{n-1/2}
  \equiv  2 \sin(\pi z)q^{1/8}\prod_{m=1}^{\infty}
    (1-q^m)(1-yq^m)(1-y^{-1}q^m), \nn [-10pt]
   & \\[-5pt]
 & \dsp \th_2(\tau,z)=\sum_{n=-\infty}^{\infty} q^{(n-1/2)^2/2} y^{n-1/2}
  \equiv 2 \cos(\pi z)q^{1/8}\prod_{m=1}^{\infty}
    (1-q^m)(1+yq^m)(1+y^{-1}q^m), \\
 & \dsp \th_3(\tau,z)=\sum_{n=-\infty}^{\infty} q^{n^2/2} y^{n}
  \equiv \prod_{m=1}^{\infty}
    (1-q^m)(1+yq^{m-1/2})(1+y^{-1}q^{m-1/2}),  
\\
 &  \dsp \th_4(\tau,z)=\sum_{n=-\infty}^{\infty}(-1)^n q^{n^2/2} y^{n}
  \equiv \prod_{m=1}^{\infty}
    (1-q^m)(1-yq^{m-1/2})(1-y^{-1}q^{m-1/2}) . 
 \end{align}
\begin{eqnarray}
 \Th{m}{k}(\tau,z)&=&\sum_{n=-\infty}^{\infty}
 q^{k(n+\frac{m}{2k})^2}y^{k(n+\frac{m}{2k})} ,
\\
\eta(\tau)  &=& q^{1/24}\prod_{n=1}^{\infty}(1-q^n).
 \end{eqnarray}
 Here, we have set $q:= e^{2\pi i \tau}$, $y:=e^{2\pi i z}$.

 We 
 use abbreviations, $\th_i (\tau) \equiv \th_i(\tau, 0)$
 ($\th_1(\tau)\equiv 0$), 
$\Th{m}{k}(\tau) \equiv \Th{m}{k}(\tau,0)$.
We also set
\begin{eqnarray}
&& Z^{(k)}[u,v](\tau) : = 
\frac{1}{\left| \eta(\tau) \right|^2}\,
\Th{u+v}{k}(\tau) \overline{\Th{-u+v}{k}(\tau)},
\label{Zuv 2}
\end{eqnarray}
which is often used in the main text.

~

\noindent
\underline{\bf Product formula of theta function:}
 \begin{equation}
 \Th{m}{k}(\tau,z)\Th{m'}{k'}(\tau,z')
 =\sum_{r\in\bsz_{k+k'}}\Th{mk'-m'k+2kk'r}{kk'(k+k')}(\tau,u)
 \Th{m+m'+2kr}{k+k'}(\tau,v),
 \label{product}
 \end{equation} 
 where we set 
 $\dsp u= \frac{z-z'}{k+k'}$, $\dsp v=\frac{kz+k'z'}{k+k'}$.

~

\noindent
\underline{\bf Poisson resummation formula:}
\begin{eqnarray}
&& \sum_{n\in\bsz}\exp\left(-\pi \al (n+a)^2+2\pi i b (n+a)\right)
=\frac{1}{\sqrt{\al}}\sum_{m\in\bsz}\exp
\left(-\frac{\pi(m-b)^2}{\al}+2\pi i m a\right), 
\nn
&& \hspace{10cm}
(\al >0 , ~ a,b \in \br).
\label{PR formula}
\end{eqnarray}

~


\noindent
\underline{\bf Modular invariant `inner product':}
\begin{equation}
 \langle \al, \beta \rangle_{\tau} := \frac{1}{\tau_2} \Im (\al \bar{\beta}).
 \label{<> 2}
\end{equation}
For example, when $\al \equiv x_1 \tau + x_2, ~ \beta \equiv y_1 \tau + y_2$, one obtains 
$\langle \al, \beta \rangle_{\tau}  = x_1 y_2- x_2 y_1$.
%
The inner product $\langle ~ , ~ \rangle_{\tau}$ is modular invariant, that is,
\begin{equation}
\langle \al, \beta \rangle_{\tau+1} = \langle \al , \beta \rangle_{\tau},
\hspace{1cm} \left\langle \frac{\al}{\tau}, \frac{\beta}{\tau} \right\rangle_{-\frac{1}{\tau}} 
= \langle \al , \beta \rangle_{\tau}.
\end{equation}
We also note 
\begin{equation}
\langle \al , \beta \rangle_{\ell \tau} = \frac{1}{\ell} \langle \al , \beta \rangle_{\tau}, \hspace{1cm}
(\any \ell \neq 0).
\end{equation}
We often use the abbreviation  
$\langle ~, ~ \rangle \equiv \langle ~, ~ \rangle_{\tau}$ in 
the cases with no fear of confusion.

~


\section*{Appendix B: ~ Diagonal Part Operator}

\setcounter{equation}{0}
\def\theequation{B.\arabic{equation}}


~

In this appendix we present the 
definition of the `diagonal part operator'
$\bD_{[\la]}$ ($[\la] \in \La/2\La$, $\any \la \in \La - 2\La$) repeatedly used in the main text.

As a preliminary, we start with a general function $F_{(w,m)}(\tau)$ ($w,m \in \bz$) 
of the form such as  
\begin{equation}
F_{(w,m)} (\tau) = \sum_{\ell, \tell}\, \sum_{n, \tilde{n} \in \bz} \, c(\ell,\tell, n, \tilde{n} ; w,m) \, 
q^{\varDelta(\ell)+n}\, \overline{q^{\widetilde{\varDelta}(\tell) + \tilde{n}}},
\end{equation}
that  possesses  the following properties, 
\begin{eqnarray}
&& \hspace{-1cm}
 F_{(-w,-m)}(\tau) = F_{(w,m)}(\tau), \hspace{1cm} 
F_{(w,m)}(\tau+1) = F_{(w, w+m)}(\tau), \hspace{1cm} F_{(w,m)}\left(-\frac{1}{\tau}\right) 
= F_{(m,-w)}(\tau).
\nn
&&
\label{assumption Fwm}
\end{eqnarray}  
We identify the `partition function' whose diagonal part is of our interest as 
\begin{equation}
\bZ_{(w,m)}(\tau) \equiv F_{(w,m)}(\tau)^2,
\label{def Zwm}
\end{equation}
and regard the two factors $F_{(w,m)}(\tau)$ as the contributions from $\mbox{CFT}_1$ 
and $\mbox{CFT}_2$.%
 \footnote
   {Of course, a natural extension of the argument given here would be 
the diagonal part operator acting on a more general partition function such as 
$ 
\bZ_{(w,m)}(\tau) \equiv F^{(1)}_{(w,m)}(\tau)\,  F^{(2)}_{(w,m)}(\tau),
\ 
F^{(1)}_{(w,m)}(\tau) \neq F^{(2)}_{(w,m)}(\tau).
$ 
In this paper, however, it is sufficient to restrict to the simple case \eqn{def Zwm}.
}

Then, we define the `diagonal part operator'  
$\bD_{[\la]}$ ($\any \la \in \La - 2\La$, $[\la] \in \La/2\La$
denotes the representative of $\la$) by
\begin{equation}
\bD_{[w\tau+m] } \cdot \bZ_{(w,m)}(\tau)  :=
\left\{
\begin{array}{ll}
  F_{(\frac{w}{2}, m)}(2\tau), & ~~  (w\in 2\bz, ~ m \in 2\bz+1),
\\
 F_{(w, \frac{m}{2})}\left(\frac{\tau}{2} \right), & ~~  (w\in 2\bz+1, ~ m \in 2\bz),
\\
 F_{(w, \frac{m-w}{2})}\left(\frac{\tau+1}{2} \right), & ~~  (w\in 2\bz+1, ~ m \in 2\bz+1).
\end{array}
\right.
\label{def bDwm}
\end{equation}
Alternatively, if adopting the notation 
such as $F_{w\tau+m}(\tau) \equiv F_{(w,m)}(\tau)$, 
$\bZ_{w\tau+m}(\tau) \equiv \bZ_{(w,m)}(\tau)$, 
we can rewrite it as 
\begin{equation}
\bD_{[w\tau+m] } \cdot \bZ_{w\tau+m}(\tau)   :=
\left\{
\begin{array}{ll}
  F_{\frac{w}{2}\cdot 2\tau+ m}(2\tau) \equiv F_{w\tau+m}(2\tau), & ~~  (w\in 2\bz, ~ m \in 2\bz+1),
\\
 F_{w\frac{\tau}{2}+ \frac{m}{2}}\left(\frac{\tau}{2} \right)
\equiv F_{\frac{1}{2}(w\tau+m)} \left(\frac{\tau}{2}\right), & ~~  (w\in 2\bz+1, ~ m \in 2\bz),
\\
 F_{w\frac{\tau+1}{2} + \frac{m-w}{2}} \left(\frac{\tau+1}{2} \right)
 \equiv F_{\frac{1}{2}(w\tau+m)} \left(\frac{\tau+1}{2} \right).
& ~~  (w\in 2\bz+1, ~ m \in 2\bz+1).
\end{array}
\right.
\label{def bDwm 2}
\end{equation}

For $\any \la = m \in  2\bz+1$,  
$\bD_{[m]} \cdot \bZ_{m} (\tau) $ is literally interpreted as the diagonal part of the partition function 
$\bZ_{m} (\tau)$,
which means 
\begin{eqnarray}
&& \bD_{[m]} \cdot \bZ_{m} (\tau) = F_{(0,m)}(2\tau)
 \equiv  
 \sum_{\ell, \tell}\, \sum_{n, \tilde{n} \in \bz} \, c(\ell,\tell, n, \tilde{n} ; 0,m) \, 
q^{2(\varDelta(\ell)+n)}\, \overline{q^{2(\widetilde{\varDelta}(\tell) + \tilde{n})}}.
\end{eqnarray}
Furthermore, the function $\bD_{[\la]} \cdot \bZ_{\la} (\tau) $
possesses the expected modularity; namely, the identities,
\begin{equation}
\bD_{[\la]} \cdot \bZ_{\la} (\tau + 1) = \bD_{[\la]} \cdot \bZ_{\la} (\tau ),
\hspace{1cm}
\bD_{[\la/\tau]} \cdot \bZ_{\la/\tau} \left(-\frac{1}{\tau}\right)  = \bD_{[\la]} \cdot \bZ_{\la} (\tau ),
\label{modularity bD}
\end{equation}
are satisfied.
In fact, almost all of the  identities are obvious from the definition \eqn{def bDwm}, 
and it is only non-trivial to prove 
\begin{equation}
\bD_{[\la/\tau]} \cdot \bZ_{\la/\tau} \left(-\frac{1}{\tau}\right)  = \bD_{[\la]} \cdot \bZ_{\la} (\tau ),
\hspace{1cm} (\any \la \in (2\bz+1) \tau + (2\bz+1)).
\end{equation}  
This is equivalent to 
the identity, 
\begin{equation}
F_{(w, \frac{m-w}{2})} \left(\frac{-\frac{1}{\tau} + 1}{2}\right) 
= F_{(m, -\frac{w+m}{2})} \left(\frac{\tau+1}{2} \right),
\hspace{1cm} (\any w,m \in 2\bz+1),
\end{equation}
which follows from \eqn{assumption Fwm} as is easily checked. 


~


\end{document}